# Basics of RF electronics

*A. Gallo*
INFN LNF

**Abstract**

RF electronics deals with the generation, acquisition and manipulation of high-frequency signals. In particle accelerators signals of this kind are abundant, especially in the RF and beam diagnostics systems. In modern machines the complexity of the electronics assemblies dedicated to RF manipulation, beam diagnostics, and feedbacks is continuously increasing, following the demands for improvement of accelerator performance. However, these systems, and in particular their front-ends and back-ends, still rely on well-established basic hardware components and techniques, while down-converted and acquired signals are digitally processed exploiting the rapidly growing computational capability offered by the available technology. This lecture reviews the operational principles of the basic building blocks used for the treatment of high-frequency signals. Devices such as mixers, phase and amplitude detectors, modulators, filters, switches, directional couplers, oscillators, amplifiers, attenuators, and others are described in terms of equivalent circuits, scattering matrices, transfer functions; typical performance of commercially available models is presented. Owing to the breadth of the subject, this review is necessarily synthetic and non-exhaustive. Readers interested in the architecture of complete systems making use of the described components and devoted to generation and manipulation of the signals driving RF power plants and cavities may refer to the CAS lectures on Low-Level RF.

## 1 Introduction

The low-level control of RF signals (LLRF) in particle accelerators is a very important continuously evolving topic since it impacts directly on the characteristics of the beam. The performance of LLRF systems has been pushed forward by the continuous growth of the required beam quality, in terms of stability, intensity, and synchronization. The complexity of LLRF systems architecture has consequently grown toward large digital systems making use of state-of-the-art signal control techniques and components.

While general LLRF is covered in specific lectures of this course, here a review of the basics of RF electronics is presented. In fact, in spite of the growing complexity of the complete systems, most of the building blocks used for RF signal manipulation (signal generation, frequency up/down conversion, amplitude and phase modulation/demodulation, filtering, matching, splitting and combining, switching/multiplexing, etc.) are still based on well-established operational principles developed in the last six decades mainly for radar and communication electronics.

The main building blocks used for RF control and beam diagnostics are summarized and analysed in the following, together with their operational principles and typical performances of commercially available models. Because of the breadth of the subject, the review cannot pretend to be exhaustive, but more realistically it is intended to be a reminder of the electronics concepts underlying the operation of the most commonly used device, from the simplest (attenuator, transformers, etc.) to the more complex ones (such as oscillators in phase locked loop configuration). A special topic such as the modulation transfer functions that have to be taken into account whenever an amplitude and/or phase modulated signal is fed into a resonant cavity has been also covered. This can be considered as an inherent extra-filtering acting on the RF drive signal which may cause distortion and/or coupling between the signal baseband components, especially in the heavy beam loading regime, and that requires special care to preserve LLRF effectiveness and stability.

The style adopted for the device description privileges the presentation of the component schematics, of the functional equations and scattering matrix, and of a list of the most commonly used specifications to qualify the performances of commercial devices. In this respect, the present review is also intended as a 'glossary' to support and orient interested students to understand the ultimate performance of real devices, providing them also with hardware selection criteria.

## 2 Fixed attenuators

Fixed attenuators are extremely simple components, widely used in RF electronics to set the proper signal level in the various circuit branches. Proper level setting is necessary to match the instrumentation dynamic range and to avoid circuit overload and damage.

Attenuators can also be used as matching pads connecting lines of different impedances. In general, the insertion of attenuators in front of mismatched loads reduces the voltage standing wave ratio (VSWR) seen at the source side.

Fixed attenuators are passive, two-port devices generally made by a network of resistors with a very broadband frequency response (dc ÷ many GHz, typically). They are designed to provide both the required attenuation and matching of the input/output lines, which might have different characteristic impedances. The attenuation ΔdB, expressed in dB units, and the linear transmission coefficient α are defined as

$$\Delta dB = 10 \cdot \log(P_{in}/P_{out}); \quad \alpha = \sqrt{P_{out}/P_{in}} = 10^{-(\Delta dB/20)}. \tag{1}$$

The associated scattering matrix is

$$S = \begin{pmatrix} 0 & \alpha \\ \alpha & 0 \end{pmatrix}. \tag{2}$$

It is important to note that in order to match unequal input/output line impedances a minimum attenuation is required, according to (case $Z_{0_{out}} \geq Z_{0_{in}}$)

$$\alpha_{max} = \sqrt{Z_{0_{out}}/Z_{0_{in}}} - \sqrt{Z_{0_{out}}/Z_{0_{in}} - 1} \Rightarrow \Delta dB_{min} = 20 \cdot \log(1/\alpha_{max}). \tag{3}$$

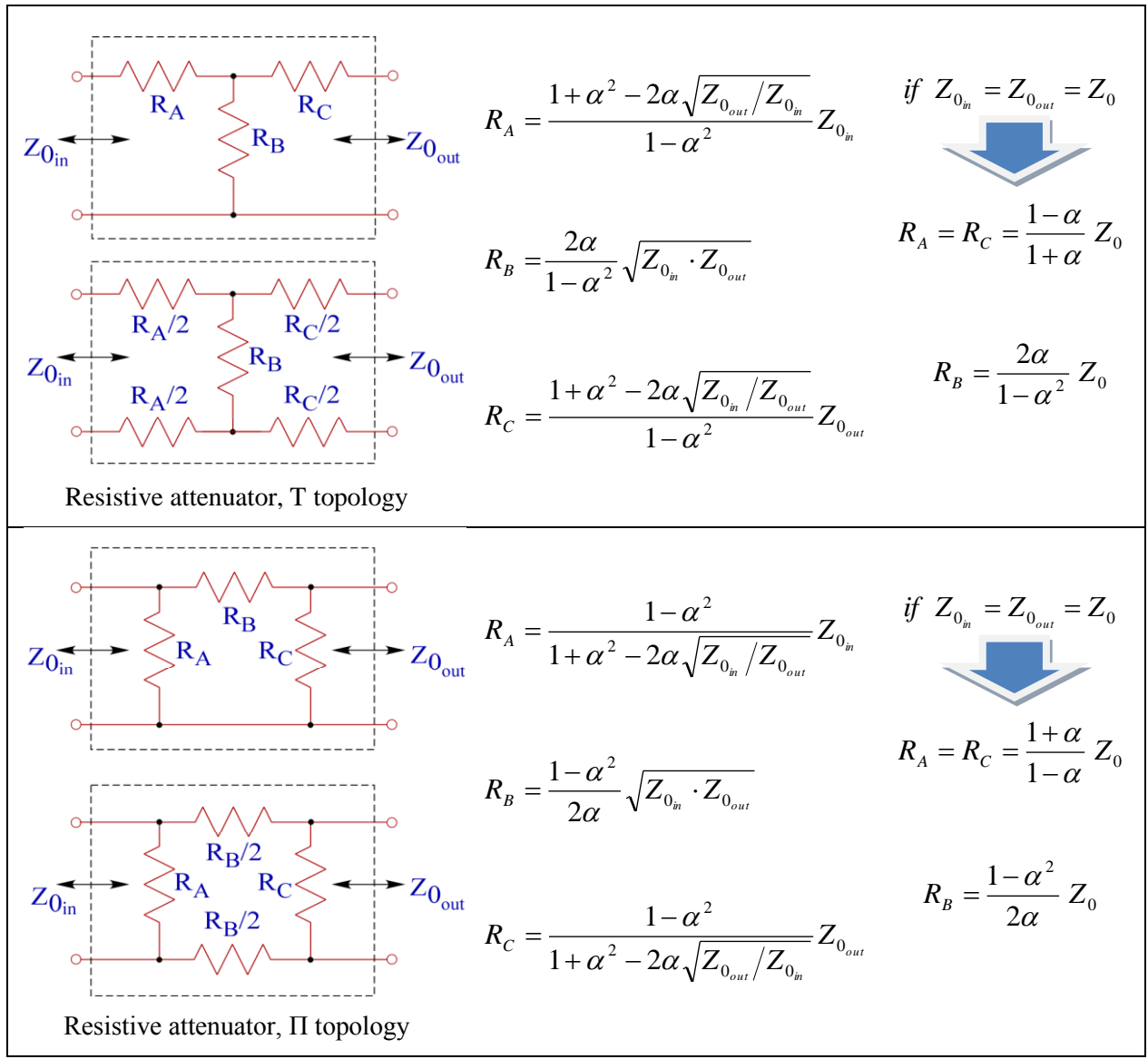

**Fig. 1:** Resistive attenuators of T and Π topologies, balanced and unbalanced configurations

Fixed attenuators are usually characterized by the following parameters:

- attenuation ΔdB;
- max. average power rate;
- max. peak power rate;
- frequency range;
- attenuation flatness over the specified frequency range;
- VSWR, size and weight, performance over the given temperature range.

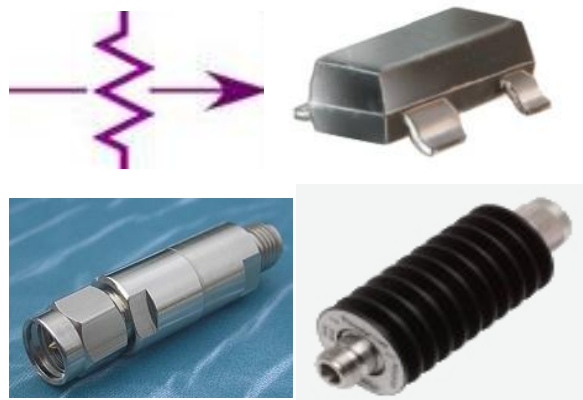

**Fig. 2:** Images of commercial RF attenuators

Fixed attenuators are available in a huge variety of packages, power ratings (up to ≈1 kW), frequency ranges (to > 18 GHz), any attenuation value, and all standard impedances used in communication electronics.

## 3  Signal amplifiers

Low-level RF amplifiers are used to increase the signal level whenever it is required for proper treatment and/or manipulation. Being a very wide subject, it can only be mentioned here. The most relevant standard specifications used to qualify the performance of small signal amplifiers are:

- **Frequency range**
  From DC to > 10 GHz, multi-decades covered by a single device.

- **Output level**
  Maximum power at the amplifier output.

- **Gain and gain flatness**
  Ratio between the output (non-saturated) and input levels, typically expressed in dB. The flatness is defined as half of the gain variation over the entire specified frequency band.

- **1 dB compression point**
  Output level corresponding to a 1 dB reduced gain because of the incipient device saturation.

- **Noise figure**
  Ratio between the input and output signal-to-noise ratios assuming an input unilateral spectral noise power density $dP_{in}/df = kT$, ($k$ = Boltzmann constant, $T$ =290 K). With $G$ its power gain, the amplifier generates an extra output spectral noise $d(\Delta P_{n_{out}})/df = G(NF-1)kT$.

- **Dynamic range**
  Potential excursion of the output level, upper-limited by compression/saturation and lower-limited by the noise power integrated over the application frequency band.

- **Two-tone third-order intercept point**
  Measures the amplifier linearity. If fed with two-tones (two equal-amplitude signals of frequencies $f_1$ and $f_2$) the amplifier generates intermodulation products at $m_1 f_1 \pm m_2 f_2$ frequencies. The amplitudes of 3$^{rd}$ order product ($2f_1 - f_2$, $2f_2 - f_1$) grow with the 3$^{rd}$ power of the input signals so that an input level corresponding to equal fundamental and 3$^{rd}$ order product amplitudes can be extrapolated (usually lying beyond the amplifier's dynamic range).

- **Input/output VSWR or return loss**
  Measure of the input/output matching characteristics of the amplifier.

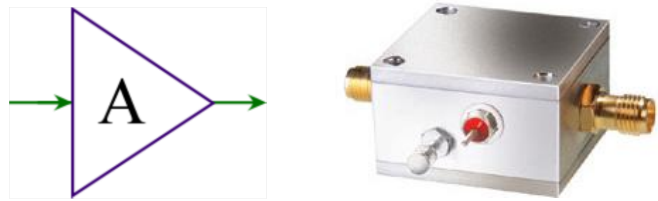

**Fig. 3:** RF amplifier symbol and image of a commercial model

Nowadays almost only solid-state technology (silicon or GaAs semiconductors, BJT and FET technology) is used for low/medium power (< 10 W) applications, up to ≈ 10 GHz. Construction techniques are MIC (Microwave Integrated Circuits) and MMIC (Monolithic Microwave Integrated

Circuits). In MIC realization the transistor and its capacitance and resistors are soldered on microstrip lines lying on a proper substrate; MMIC are completely integrated circuits where all components (the transistors and their ancillaries) are fabricated on a common substrate.

Small-signal amplifiers generally operate in class 'A' since power efficiency is not an issue in this context.

## 4  RF transformers

Transformers are widely used in RF electronics. Their operating principle is a direct consequence of the Faraday–Neumann–Lenz law of electromagnetic induction. They are very effective to
- match lines of different impedance with negligible insertion loss;
- de-couple ground while transmitting RF signals;
- connect balanced and unbalanced circuits (bal-un).

RF transformers can also be found embedded in a number of other devices (splitters/combiners, mixers, amplifiers, etc.).

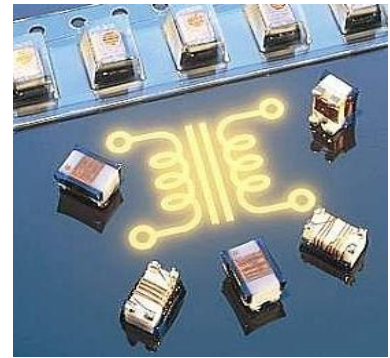

**Fig. 4:** RF transformers

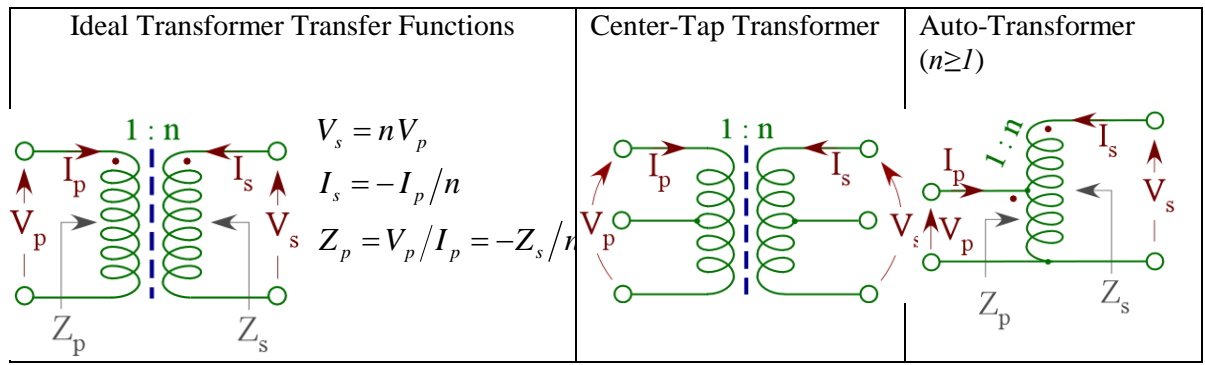

**Fig. 5:** Different types of ideal transformers

Together with transform ratio $n$ and connection topology, real transformers are characterized by operating bandwidth, insertion loss, maximum power rating, etc. A sufficiently accurate circuit model for real transformers is represented in Fig. 6. The device lower cutoff frequency is due to the winding's active inductance $L_{act}$, while the high-frequency cutoff is dominated by the inter-winding and intra-winding capacitances $C_{p-p}$, $C_{s-s}$ and $C_{p-s}$. In-band insertion loss is due to the magnetic core dissipation and to the winding's ohmic losses, accounted by the resistances $R_{loss}$, $R_p$ and $R_s$.

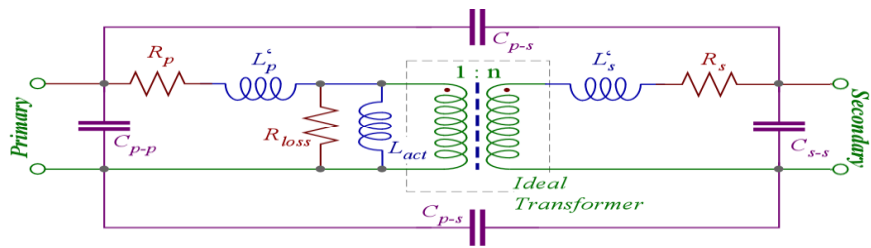

**Fig. 6:** Circuit model of a real transformer

## 5    Power splitters/combiners

Power splitters/combiners are *N*-port devices used to divide a signal into $N-1$ equal copies ($N$ = any number, typically $N = 2^k + 1$), or to make a vector sum of $N-1$ different signals.

Basic characteristics of this kind of device are:

- number of channels *N*;
- operating frequency range;
- maximum power ratings;
- splitting technique (reactive or resistive);
- insertion loss over the nominal $10\,Log[N-1]$;
- isolation between channels;
- Phase and amplitude unbalance among output channels;
- etc.

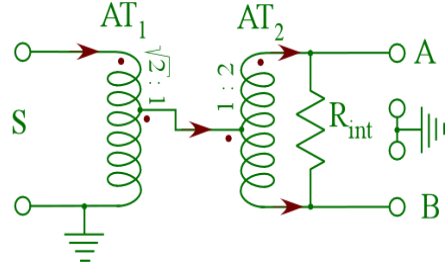

**Fig. 7:** A two-way reactive splitter/combiner based on a double tapped auto-transformer provides impedance matching at all ports and isolation between A and B channels

Ideally, the signal at the common port is split into $N-1$ copies, and the power into any output channel is $P_{out} = P_{in}/(N-1)$. Assuming port 1 as the splitter/combiner common port, the scattering matrix of an ideal *N*-port splitter/combiner with perfectly isolated channels is

$$S = \begin{pmatrix} 0 & 1/\sqrt{N-1} & \dots & 1/\sqrt{N-1} \\ 1/\sqrt{N-1} & 0 & \dots & 0 \\ \dots & \dots & \dots & \dots \\ 1/\sqrt{N-1} & 0 & \dots & 0 \end{pmatrix}. \qquad (4)$$

A very simple network implementing the functionalities required for a 3-port combiner/splitter is shown in Fig. 8. Isolation between ports A and B is obtained by a proper choice of the $R_{int}$ value ($R_{int} = 2Z_0$) in the second auto-transformer, while the first transformer is needed for impedance matching [see Fig. 8 and Eq. (5)].

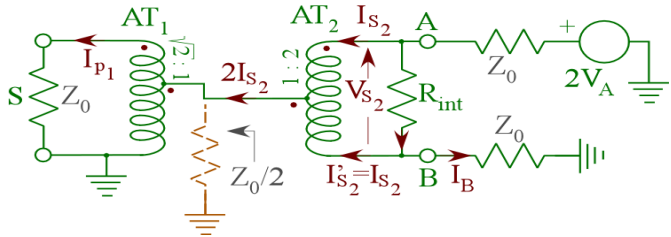

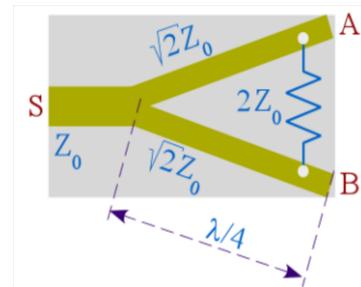

**Fig. 8:** Circuital analysis of a transformer-based combiner/splitter

**Fig. 9:** Microstrip (Wilkinson) combiner/splitter

$$V_{p_2} = V_{s_2}/2; \quad I_{p_2} = 2I_{s_2} \quad \Rightarrow AT_2 \text{ Transformer equations}$$

$$I_B = V_{s_2}/R_{int} - I_{s_2} = I_{s_2}\left(2Z_0/R_{int} - 1\right) \underset{\underset{R_{int} = 2Z_0}{if}}{=} 0 \quad \Rightarrow \text{Kirchhoff current law @ node B} \qquad (5)$$

Microstrips can be used instead of transformers for high-frequency, octave-band devices, as shown in Fig. 9 (Wilkinson power divider).

Splitter/combiners can be also resistive, consisting in a 'star' connection of equal resistors (Fig. 10). The frequency response can be much wider (extending from DC) and flatter in this case, at the expense of a larger insertion loss and no isolation (all ports equally coupled).

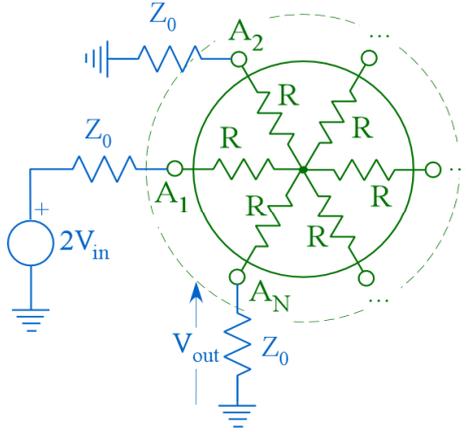

$$Matching \Rightarrow R = \frac{N-2}{N} Z_0$$

$$Transmission \Rightarrow \frac{P_{out}}{P_{in}} = \left(\frac{V_{out}}{V_{in}}\right)^2 = \frac{1}{(N-1)^2}$$

**Fig. 10:** *N*-port resistive power splitter/combiner

There are no isolated ports in this kind of splitter/combiner, whose scattering matrix is

$$S = \begin{pmatrix} 0 & 1/(N-1) & \ldots & 1/(N-1) \\ 1/(N-1) & 0 & \ldots & 1/(N-1) \\ \ldots & \ldots & \ldots & \ldots \\ 1/(N-1) & 1/(N-1) & \ldots & 0 \end{pmatrix} . \tag{6}$$

As the insertion loss grows linearly with the number of ports, practical use is restricted to 3-port devices.

## 6   Hybrid junctions/directional couplers

Hybrid junctions and directional couplers are 4-port passive devices based on the same operational principles but with different coupling levels between ports. The two classes of devices are used for different purposes. Hybrids are used whenever it is necessary to split/combine signals with specific phase relations, while directional couplers (i.e., hybrids with coupling coefficient different from -3 dB) are used to sample forward and/or reflected waves propagating along a line.

Symbolic representation and scattering matrices of directional couplers, 90° hybrids, and 180° hybrids are shown in Figs. 11, 12 and 13.

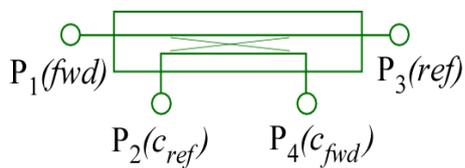

$$S = \begin{pmatrix} 0 & 0 & -j\sqrt{1-c^2} & c \\ 0 & 0 & c & -j\sqrt{1-c^2} \\ -j\sqrt{1-c^2} & c & 0 & 0 \\ c & -j\sqrt{1-c^2} & 0 & 0 \end{pmatrix}$$

**Fig. 11:** Directional coupler Symbolic representation and scattering matrix

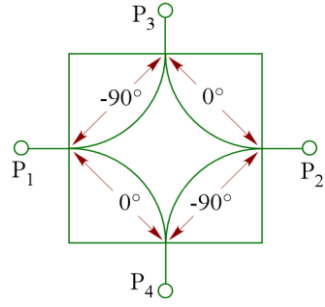

$$S = \frac{1}{\sqrt{2}} \begin{pmatrix} 0 & 0 & -j & 1 \\ 0 & 0 & 1 & -j \\ -j & 1 & 0 & 0 \\ 1 & -j & 0 & 0 \end{pmatrix}$$

**Fig. 12:** 90° Hybrid
Symbolic representation and scattering matrix

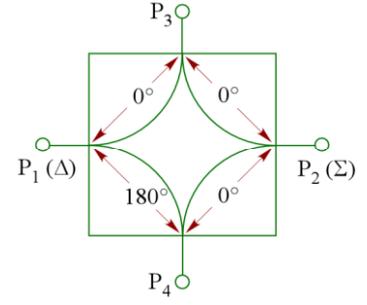

$$S = \frac{-j}{\sqrt{2}} \begin{pmatrix} 0 & 0 & 1 & -1 \\ 0 & 0 & 1 & 1 \\ 1 & 1 & 0 & 0 \\ -1 & 1 & 0 & 0 \end{pmatrix}$$

**Fig. 13:** 180° Hybrid
Symbolic representation and scattering matrix

### 6.1 Continuously coupled lines

Distributed coupling between two lines travelling close to each other is one of the possible layouts of a hybrid/coupler. The lines have a characteristic impedance $Z_0$ when travelling separately, while the 3-conductor system of the two coupled lines has even and odd excitation impedances $Z_0^+$ and $Z_0^-$, respectively.

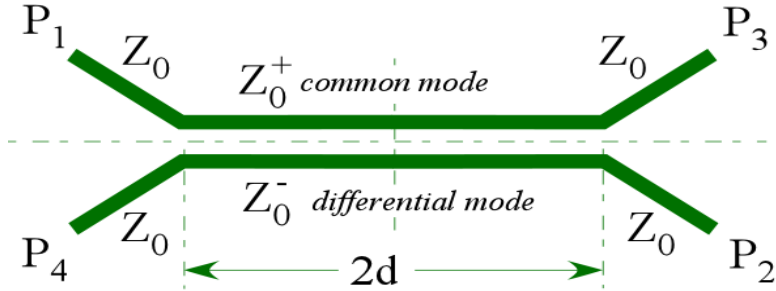

**Fig. 14:** Coupled line directional coupler

The scattering matrix can be worked out by exploiting the 4-fold symmetry of the network. With $\beta^\pm$ the propagation constants of even and odd modes, we get

$$\left. \begin{array}{l} \beta^+ = \beta^- \equiv \beta;\ Z_0^+ Z_0^- = Z_0^2 \\[4pt] c = \dfrac{c_{max} \sin(2\beta d)}{\sqrt{1 - c_{max}^2 \cos^2(2\beta d)}} \\[4pt] \text{with:}\ c_{max} = \dfrac{Z_0^+ - Z_0^-}{Z_0^+ + Z_0^-} \\[4pt] \tan\varphi_c = \dfrac{\sqrt{1 - c_{max}^2}}{\tan(2\beta d)} \end{array} \right\} \Rightarrow S = -j \begin{pmatrix} 0 & 0 & \sqrt{1-c^2}\,e^{j\varphi_c} & jc\,e^{j\varphi_c} \\ 0 & 0 & jc\,e^{j\varphi_c} & \sqrt{1-c^2}\,e^{j\varphi_c} \\ \sqrt{1-c^2}\,e^{j\varphi_c} & jc\,e^{j\varphi_c} & 0 & 0 \\ jc\,e^{j\varphi_c} & \sqrt{1-c^2}\,e^{j\varphi_c} & 0 & 0 \end{pmatrix} . \quad (7)$$

If the length of the coupled line is exactly $2d = \lambda/4 = \pi/(2\beta)$, the scattering matrix has its simplest form:

$$\left.\begin{array}{l} 2\beta d = \pi/2 \\ \Downarrow \\ c = c_{max} = \dfrac{Z_0^+ - Z_0^-}{Z_0^+ + Z_0^-} \\ \varphi_c = 0 \end{array}\right\} \Rightarrow S = -j \begin{pmatrix} 0 & 0 & \sqrt{1-c_{max}^2} & jc_{max} \\ 0 & 0 & jc_{max} & \sqrt{1-c_{max}^2} \\ \sqrt{1-c_{max}^2} & jc_{max} & 0 & 0 \\ jc_{max} & \sqrt{1-c_{max}^2} & 0 & 0 \end{pmatrix}. \quad (8)$$

If coupling factors lower than 10 dB ($c_{max} > 0.3$) are required, broadside coupled strips can be used, as shown in Fig. 15. At $c_{max} = 1/\sqrt{2}$ (i.e., $Z_0^+ \approx 5.83\, Z_0^-$) the device scattering matrix is that of a 90° hybrid junction. Coupled wound coils on ferrite cores are used at low frequencies.

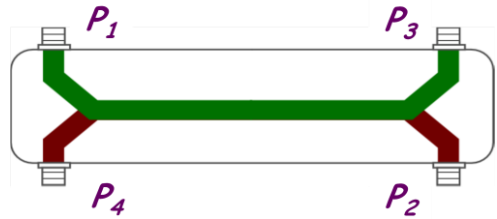

**Fig. 15:** Broadside coupled microstrips

## 6.2 Hole-coupled lines

Another possible directional coupler structure is represented by two parallel lines connected through coupling holes. Two equal holes separated by $\lambda/4$ are sufficient to produce the proper scattering matrix at a given frequency. Device bandwidth can be enlarged with multiple holes with different optimal dimensions.

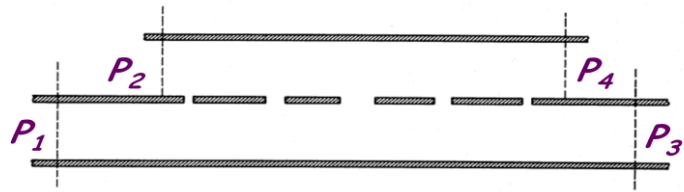

**Fig. 16:** Hole-coupled lines

This technique is mainly used in waveguide directional couplers.

## 6.3 Branch line coupler

The branch line coupler is another layout producing the desired port-to-port coupling.

The scattering matrix can be still worked out by exploiting the 4-fold symmetry of the network. Under the assumptions

I)  $\beta d_2 = \beta d_1 = \pi/2$,

II) $1/Z_1^2 - 1/Z_2^2 = 1/Z_0^2$,

the scattering matrix is that of a directional coupler:

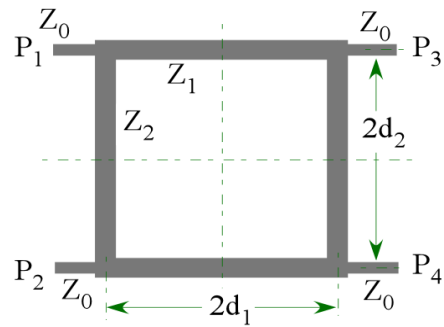

**Fig. 17:** Branch line coupler

$$S = \begin{pmatrix} 0 & 0 & -jZ_1/Z_0 & -Z_1/Z_2 \\ 0 & 0 & -Z_1/Z_2 & -jZ_1/Z_0 \\ -jZ_1/Z_0 & -Z_1/Z_2 & 0 & 0 \\ -Z_1/Z_2 & -jZ_1/Z_0 & 0 & 0 \end{pmatrix}, \quad (9)$$

which, under the conditions

III)  $Z_2 = Z_0$,

IV)  $Z_1 = Z_0/\sqrt{2}$,

becomes that of a 90° hybrid.

### 6.4  Hybrid ring

The hybrid ring (also called 'rat-race') coupler is a suitable geometry to obtain 180° hybrids.

Under the assumptions

I)  $d_{1,2} = d_{2,3} = d_{3,4} = \pi/2\beta = \lambda/4$,

II)  $d_{1,4} = 3\pi/2\beta = 3\lambda/4$,

the scattering matrix results are isolated (port 1 vs. port 2 and port 3 vs. port 4) but unmatched.

$$S = \begin{pmatrix} \dfrac{Z_1^2 - 2Z_0^2}{Z_1^2 + 2Z_0^2} & \dfrac{-2jZ_1Z_0}{Z_1^2 + 2Z_0^2} & 0 & \dfrac{2jZ_1Z_0}{Z_1^2 + 2Z_0^2} \\ \dfrac{-2jZ_1Z_0}{Z_1^2 + 2Z_0^2} & \dfrac{Z_1^2 - 2Z_0^2}{Z_1^2 + 2Z_0^2} & \dfrac{-2jZ_1Z_0}{Z_1^2 + 2Z_0^2} & 0 \\ 0 & \dfrac{-2jZ_1Z_0}{Z_1^2 + 2Z_0^2} & \dfrac{Z_1^2 - 2Z_0^2}{Z_1^2 + 2Z_0^2} & \dfrac{-2jZ_1Z_0}{Z_1^2 + 2Z_0^2} \\ \dfrac{2jZ_1Z_0}{Z_1^2 + 2Z_0^2} & 0 & \dfrac{-2jZ_1Z_0}{Z_1^2 + 2Z_0^2} & \dfrac{Z_1^2 - 2Z_0^2}{Z_1^2 + 2Z_0^2} \end{pmatrix} \quad (10)$$

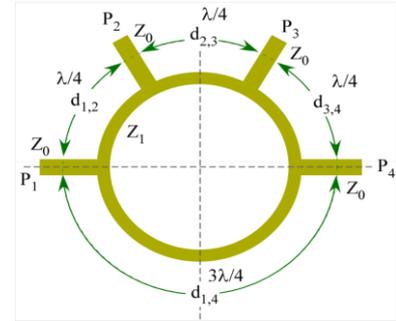

**Fig. 18:** Hybrid ring

Under the condition $Z_1 = \sqrt{2}\,Z_0$ the scattering matrix is also matched and the network behaves as an ideal 180° hybrid (or an ideal magic-T):

$$S = \dfrac{-j}{\sqrt{2}} \begin{pmatrix} 0 & 1 & 0 & -1 \\ 1 & 0 & 1 & 0 \\ 0 & 1 & 0 & 1 \\ -1 & 0 & 1 & 0 \end{pmatrix} \quad ; \quad S \otimes \begin{bmatrix} V_1 \\ 0 \\ V_3 \\ 0 \end{bmatrix} = \dfrac{-j}{\sqrt{2}} \begin{bmatrix} 0 \\ V_3 + V_1 \\ 0 \\ V_3 - V_1 \end{bmatrix}. \quad (11)$$

As evidenced in Eq. (11), it is noticeable that signals applied to a pair of uncoupled ports (1 & 3 or 2 & 4) appear as vector sum and difference at the other pair of uncoupled ports. This is an important property exploited in beam diagnostics to obtain, for instance, beam transverse position from striplines or beam position monitors.

## 6.5 Basic characteristics of a real directional coupler

Real devices cannot present perfect isolation between nominally uncoupled ports. Imperfect isolation is quantified by a parameter called directivity $d$ defined as the ratio between the nominally uncoupled and coupled elements of the real scattering matrix, as shown in Eq. (12). Directivity is often expressed in logarithmic units $D_{dB} = -20 \cdot \log |d|$.

$$S = \begin{pmatrix} 0 & c \cdot d & -j\sqrt{1-c^2} & c \\ c \cdot d & 0 & c & -j\sqrt{1-c^2} \\ -j\sqrt{1-c^2} & c & 0 & c \cdot d \\ c & -j\sqrt{1-c^2} & c \cdot d & 0 \end{pmatrix} . \tag{12}$$

The most relevant characteristics of a real directional coupler/hybrid network are

- coupling coefficient (3 dB for hybrids);
- directivity/isolation between uncoupled ports;
- operating frequency range;
- maximum power ratings;
- coupling type (holes, distributed, rings);
- insertion loss (over the nominal coupling factor);
- phase and amplitude unbalance among output channels (hybrids);
- phase and amplitude flatness over frequency;
- etc.

## 7 Filters

Filters are 2-port devices 'tailored' to obtain a specific required frequency response (the $s_{21}$ of the network). Filters are described in general as linear networks. In time domain the filter response is characterized by its Green function $H(\tau)$ so that, if excited at the input with a signal $v_i(t)$, the filter generates an output signal $v_o(t)$ given by the convolution integral

$$v_o(t) = \int_0^\infty H(\tau) v_i(t-\tau) d\tau . \tag{13}$$

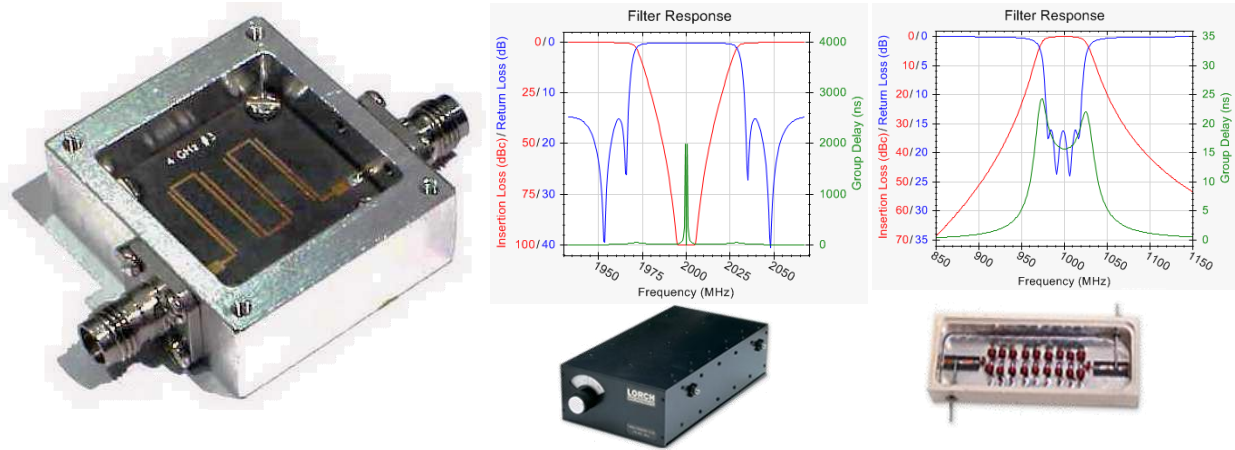

**Fig. 19:** Commercial filters of various types and different frequency response

The Green function $H(\tau)$ represents the response of the filter at a time $\tau$ after the application of an elementary Dirac stimulus $\delta(t)$. In frequency domain, input and output signals are substituted by their Fourier transforms, and the convolution integral is transformed in a linear multiplication operator according to

$$V_o(j\omega) = H(j\omega) \cdot V_o(j\omega) \quad , \tag{14}$$

where $H(j\omega)$, the Fourier transform of $H(\tau)$, is known as the filter transfer function. Typically the response $H(j\omega)$ is maximized at some bands of interest, and minimized at other frequency bands that have to be rejected.

Filters are classified on the basis of their nature, topology, dissipation, etc.

– **Analog/digital**, depending on the nature (continuous or sampled & digitized) of the input signal;

– **Lumped/distributed**, depending on the nature of the internal components (L-C cells, dielectric resonator cavities, µ-strip cells, etc.);

– **Reflective/absorbing**, depending on the path of the stopbands (reflected or internally dissipated);

– **Lowpass/highpass/bandpass/notch/comb**, depending on the profile of the frequency response.

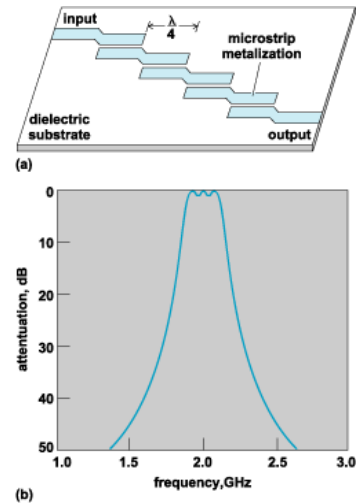

**Fig. 20:** Microstrip bandpass filter

Filters with **different response** around **transition** between pass and stop bands are available for different applications. They implement **different rational complex polynomials** in their transfer functions, the most popular ones being:

- **Bessel,** for a maximally flat group delay;

- **Butterworth,** for a maximally flat frequency response in the pass-band;

- **Gaussian,** providing a Gaussian response to a Dirac pulse and no overshoot for an input step function. The Gaussian filter also minimizes the group delay;

- **Chebyshev,** providing a steep transition with some passband (type I) or stopband (type II) ripples. They provide the closest possible response w.r.t. an ideal rectangular filter;

- **Elliptic,** providing the steepest possible transition between the pass-band and the stop-band by equalizing the ripple in both.

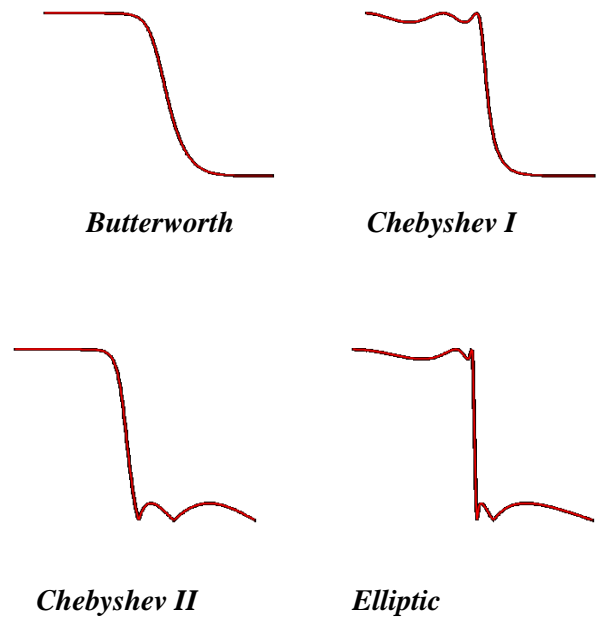

*Butterworth*   *Chebyshev I*

*Chebyshev II*   *Elliptic*

**Fig. 21:** Frequency response of different polynomial filters

The design and construction of filters is becoming more and more a specialized activity, so that 'home-made' devices are seldom used, and mainly for very specific tasks. The design phase make use of dedicated software packages through various iterative steps. In addition to the typologies already listed, other important characteristics defining filter performance are:

- **Insertion loss**, defined as the in-band signal attenuation;

- **Phase linearity/group delay**: figures of the quality of the filter phase response across the pass-band, that should present a constant negative slope to avoid distortion of the time-profile of in-band pulses. The **slope** of the **response phase** is the filter **group delay**, equal to the signal latency while travelling across the device;

- **Input/output impedance** and **VSWR**: characteristic impedance of the device and reflectivity of signals transmitted (within pass-band) and rejected (within stop-band).

## 7.1 Digital filtering

Digital filters act on sampled and digitized input signals. There are two basic filter architectures:

- **Finite Impulsive Response** (FIR), where the output *y* is a linear combination of the last *N* sampled values of the input *x*. The coefficients $h_i$ of the expansion represent the discretization of the Green function of the filter.
- **Infinite Impulsive Response** (IIR), where the output *y* is a linear combination of the last *N* and *M* sampled values of the input *x* and output *y,* respectively. IIR filters directly implement a feedback architecture, which allows the generation of sharp frequency responses with a limited number of samples.

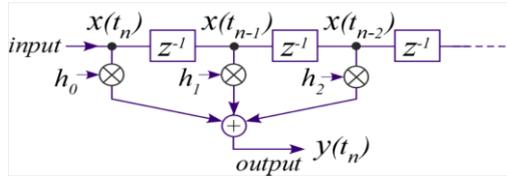

**Fig. 22:** FIR filter architecture

$$H(\tau) = \text{filter Green function} \quad y(t) = \int_0^\infty H(\tau) x(t-\tau) d\tau$$

$$y(t_n) = h_0 x(t_n) + h_1 x(t_{n-1}) + h_2 x(t_{n-2}) + \ldots = \sum_{i=0}^N h_i x_{n-i}$$

$$H(z) = \frac{Y(z)}{X(z)} = h_0 + h_1 z^{-1} + h_2 z^{-2} + \ldots = \sum_{i=0}^N h_i z^{-i}$$

(15)

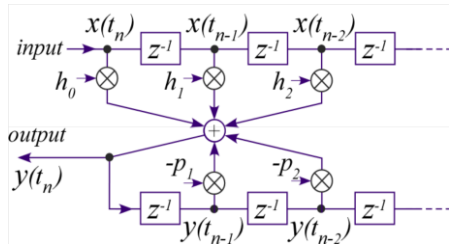

**Fig. 23:** IIR filter architecture

$$y(t_n) = h_0 x(t_n) + h_1 x(t_{n-1}) + h_2 x(t_{n-2}) + \ldots$$

$$\ldots - p_1 y(t_{n-1}) - p_2 x(t_{n-2}) + \ldots = \sum_{i=0}^N h_i x_{n-i} - \sum_{k=1}^M p_k y_{n-k}$$

$$H(z) = \frac{Y(z)}{X(z)} = \frac{h_0 + h_1 z^{-1} + h_2 z^{-2} + \ldots}{1 + p_1 z^{-1} + p_2 z^{-2} + \ldots} = \frac{\sum_{i=0}^N h_i z^{-i}}{1 + \sum_{k=1}^M p_k z^{-k}}$$

(16)

The $z$-domain transfer function $H_z(z)$ gives direct information on the filter frequency response being related to the Laplace $H_L(s)$ and Fourier $H_F(j\omega)$ transfer functions by the mathematical expressions

$$H_L(s) = H_z(z)\big|_{z=e^{sT}} ; \quad H_F(j\omega) = H_z(z)\big|_{z=e^{j\omega T}} .$$

(17)

As an example, the frequency response of a comb filter implementing the IIR architecture is shown in Fig. 24. The frequency response peaks regularly at multiples of a certain fundamental equal to 1/80 of the sampling frequency $f_0 = f_{samp}/80$. Such a filter can be used, for instance, to extract the revolution harmonics of a beam-induced signal in a pick-up in circular accelerators, and can be simply and effectively implemented in an IIR configuration with a single non-zero coefficient:

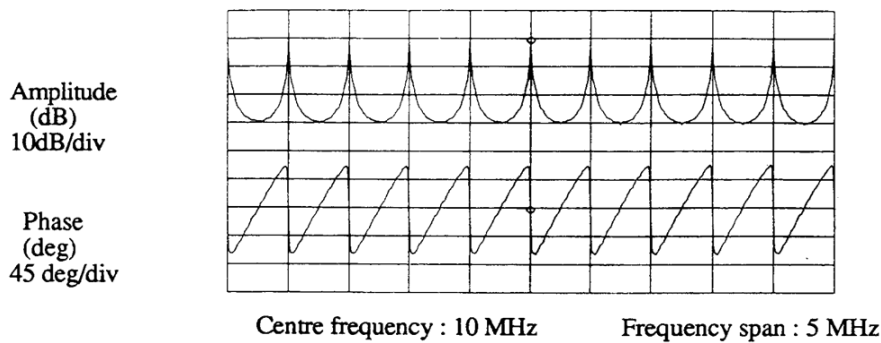

$$p_k = \begin{cases} 0, & k \neq 80 \\ 1 - 2^{-4}, & k = 80 \end{cases}$$

$$H_z(z) = \frac{1}{1 - p_k z^{-k}}$$

**Fig. 24:** Measured frequency response of a digital IIR comb filter

Differences between analog and digital filtering are quite evident. Digital filtering is a complex operation requiring many steps such as down-conversion (necessary in most cases), A-to-D conversion, digital data manipulation, D-to-A conversion, and final frequency up-conversion.
However, powerful ICs available today are capable of performing various tasks simultaneously.
On the other hand, digital filtering provides incomparable flexibility and operational adaptivity, since the transfer function can be modified and optimized in real time by simply changing the weighting coefficients. Beam feedbacks can greatly benefit this feature.

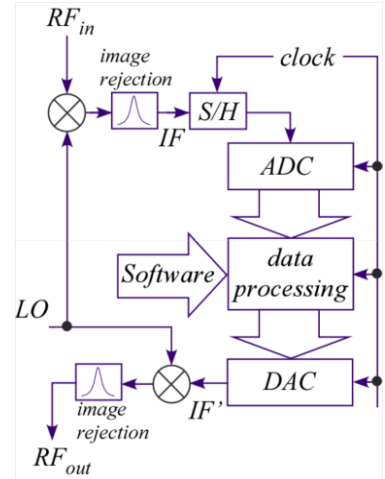

**Fig. 25:** Schematics of digital signal manipulation

### 7.2 Modulation transfer function filtering

A very peculiar filtering process that RF electronics often have to deal with is related to the so-called 'modulation transfer functions'.

RF servo-loops and feedback loops generally need to apply amplitude (AM) and/or phase (PM) modulation to the RF drive signal. The response of a resonant cavity to AM and PM excitations depends on its bandwidth and tuning relative to the carrier. If the cavity resonant frequency $\omega_r$ is tuned exactly on the RF carrier frequency $\omega_c$, left and right modulation sidebands are equally filtered, so that an input AM (PM) signal produces an AM (PM) output, and no mixing between the two modulation species occurs:

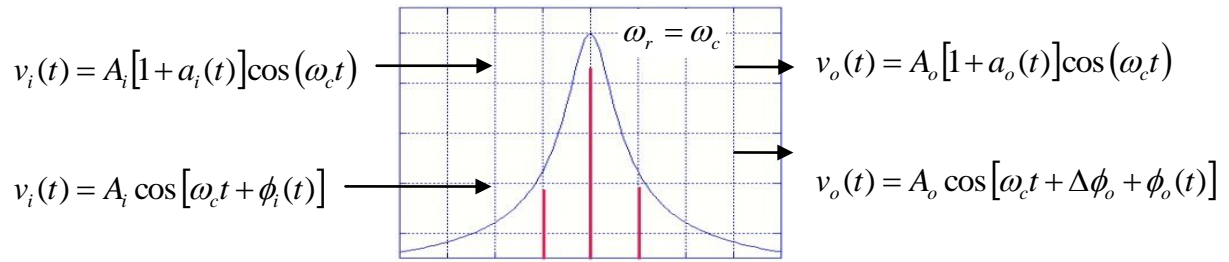

**Fig. 26:** AM/PM signals symmetrically filtered by a resonant cavity

The modulation transfer function $G(s)$ is the ratio between the Laplace transforms of the output and the input modulating functions:

$$G(s) = \frac{\hat{a}_o(s)}{\hat{a}_i(s)} = \frac{\hat{\varphi}_o(s)}{\hat{\varphi}_i(s)} = \frac{1}{1+s/\sigma} \quad \text{with} \quad \sigma = \frac{\omega_r}{2Q_L} \quad . \tag{18}$$

In the simplest case $\omega_r = \omega_c$ the cavity filters the modulating AM and PM functions with a transfer function $G(s)$ corresponding to a single-pole low-pass filter whose cut-off frequency is the cavity half-bandwidth.

A more complex situation appears when carrier and cavity resonant frequencies do not coincide. In the case $\omega_r \neq \omega_c$ left and right modulation sidebands are treated asymmetrically by the cavity transfer function, so that a single modulation type (AM or PM) at the input will result in both modulation types (AM and PM) at the output.

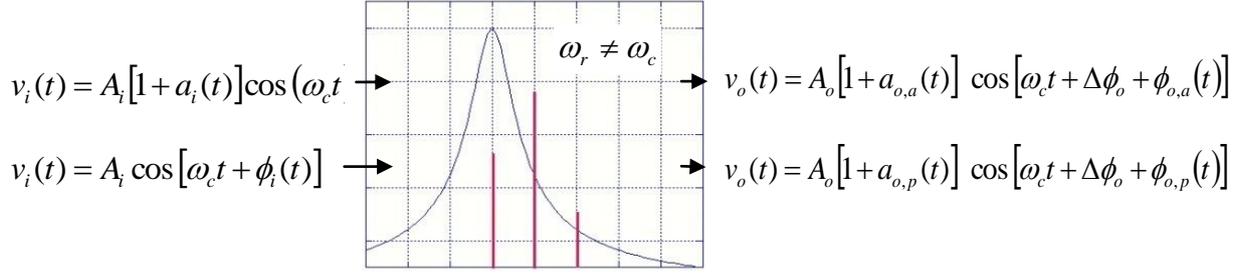

**Fig. 27:** AM/PM signals asymmetrically filtered by a resonant cavity

An off-resonance cavity produces modulation mixing, resulting in four modulation transfer functions, defined as the ratio of the Laplace transforms of the output and input modulating functions:

$$G_{aa}(s) = \frac{\hat{a}_{o,a}(s)}{\hat{a}_i(s)}; \quad G_{pp}(s) = \frac{\hat{\varphi}_{o,p}(s)}{\hat{\varphi}_i(s)}; \quad G_{ap}(s) = \frac{\hat{\varphi}_{o,a}(s)}{\hat{a}_i(s)}; \quad G_{pa}(s) = \frac{\hat{a}_{o,p}(s)}{\hat{\varphi}_i(s)} \quad . \tag{19}$$

It may be demonstrated that for a generic linear network defined by a linear transfer function $A(s)$, direct and cross modulation transfer functions are given by

$$G_{pp}(s) = G_{aa}(s) = \frac{1}{2}\left[\frac{A(s+j\omega_c)}{A(j\omega_c)} + \frac{A(s-j\omega_c)}{A(-j\omega_c)}\right] ,$$

$$G_{ap}(s) = -G_{pa}(s) = \frac{1}{2j}\left[\frac{A(s+j\omega_c)}{A(j\omega_c)} - \frac{A(s-j\omega_c)}{A(-j\omega_c)}\right] . \tag{20}$$

In the case we are considering, the linear network is a resonant cavity so that the transfer function $A(s) = A_{cav}(s)$ is

$$A_{cav}(s) = A_0 \frac{2\sigma s}{s^2 + 2\sigma s + \omega_r^2} \quad \text{with} \quad \omega_r \approx \omega_c + \sigma \tan \varphi_z \quad , \tag{21}$$

where $\phi_z$ is the cavity tuning angle, i.e., the phase of the cavity transfer function at the carrier frequency $\omega_c$. Finally, substituting Eq. (21) into Eq. (20), one gets

$$G_{pp}(s) = G_{aa}(s) = \frac{\sigma s + \sigma^2 \left(1 + \tan^2 \varphi_z\right)}{s^2 + 2\sigma s + \sigma^2 \left(1 + \tan^2 \varphi_z\right)} ,$$

$$G_{ap}(s) = -G_{pa}(s) = -\frac{\sigma \tan \varphi_z \, s}{s^2 + 2\sigma s + \sigma^2 \left(1 + \tan^2 \varphi_z\right)} . \tag{22}$$

The general form of the modulation transfer functions features two poles (in general a complex conjugate pair) and one zero. It is easy to see how cross modulation vanishes if the cavity is perfectly tuned ($\phi_z = 0$), while direct modulation terms reduce to the expression of Eq. (18). The magnitude of direct and cross modulation transfer functions for various cavity detuning values is plotted in Figs. 28 and 29.

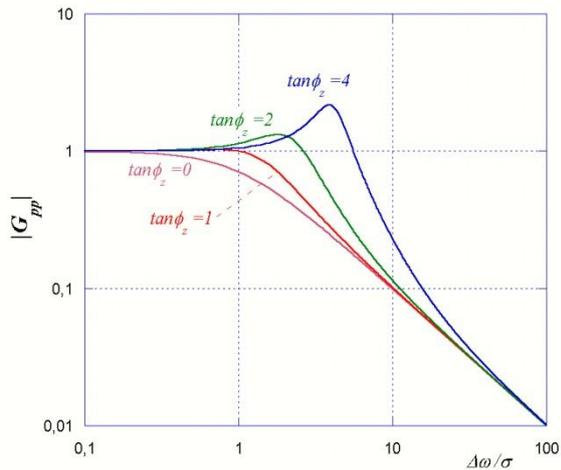 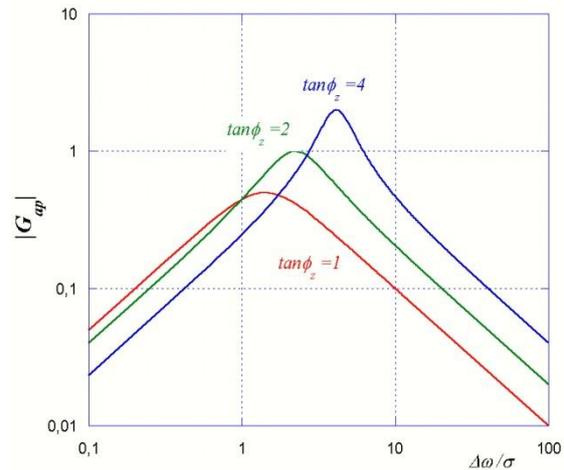

**Fig. 28:** Magnitude of direct modulation transfer function  
**Fig. 29:** Magnitude of cross modulation transfer function

      The role of the modulation transfer functions is essential in the RF system for circular accelerators under heavy beam loading. Compensation of the reactive beam loading requires a substantial detuning of the cavity accelerating mode with respect to the RF carrier. The beam phase depends on the accelerating voltage phase through the so-called 'beam transfer function' $B(s)$ which contains the synchrotron dynamics, while the accelerating voltage amplitude and phase depend on the beam intensity and phase and on the RF power source amplitude and phase through the modulation transfer functions. Because of the cavity detuning, cross modulation terms are not negligible. The whole generator–cavity–beam linear system can be graphically represented in a diagram called the Pedersen model represented in Fig. 30. The modulation transfer functions vary with the stored current and definitely couple the servo-loops and the beam loops implemented around the system.

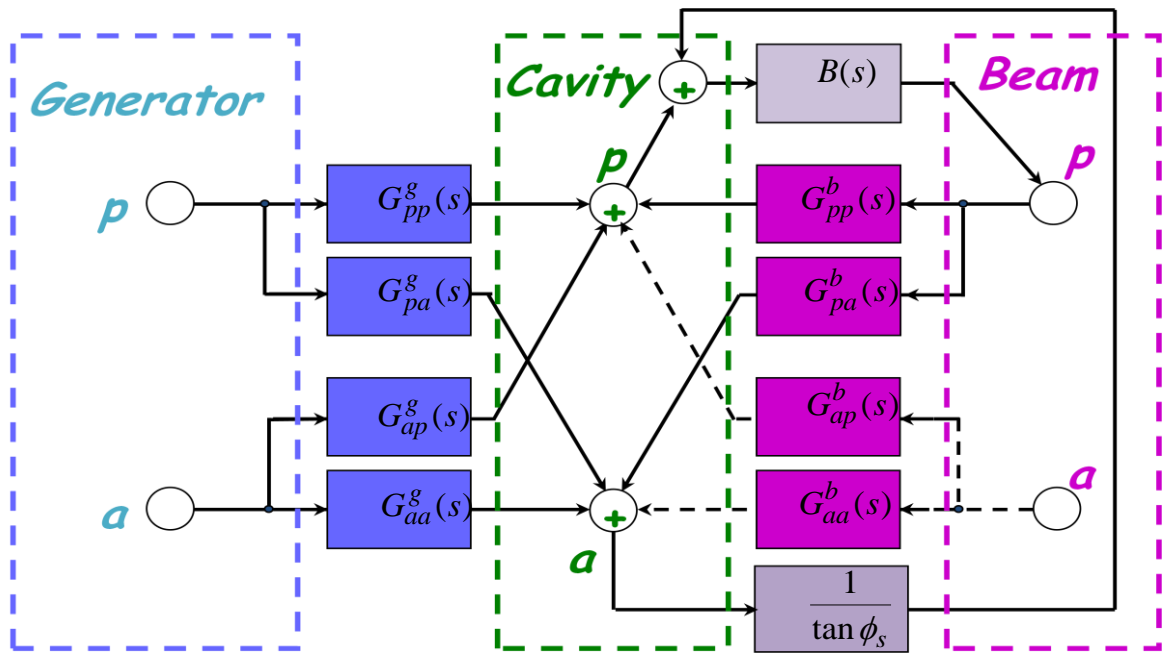

**Fig. 30:** Pedersen model of a generator–cavity–beam system

## 8  Frequency mixers

Frequency mixers are non-linear, generally passive devices used in a huge variety of RF applications. Basically, a mixer is used to perform the frequency translation of the spectrum of an RF signal to be manipulated. The spectrum shift is obtained from an analog multiplication between the RF signal and a Local Oscillator (LO).

Let's consider processing a signal $V_{RF}(t)$ through analog multiplication with a reference sine wave generally indicated as 'local oscillator' $V_{LO}\cos(\omega_{LO}t)$. With $k$ the analog multiplier conversion factor, the output signal $V_{IF}(t)$ is given by

$$V_{IF}(t) = kV_{RF}(t)V_{LO}\cos(\omega_{LO}t) \quad . \tag{23}$$

The Fourier transforms of the input and output signals $V_{RF}(t)$ and $V_{IF}(t)$ are given by

$$\tilde{V}_{RF}(\omega) = \tilde{F}[V_{RF}(t)] = \frac{1}{\sqrt{2\pi}}\int_{-\infty}^{+\infty} V_{RF}(t)e^{-j\omega t}\,dt \quad,$$

$$\tilde{V}_{IF}(\omega) = \frac{kV_{LO}}{2}\frac{1}{\sqrt{2\pi}}\int_{-\infty}^{+\infty} V_{RF}(t)e^{-j\omega t}\left(e^{j\omega_{LO}t} + e^{j\omega_{LO}t}\right)dt = \frac{kV_{LO}}{2}\left[\tilde{V}_{RF}(\omega-\omega_{LO}) + \tilde{V}_{RF}(\omega+\omega_{LO})\right] \quad . \tag{24}$$

Clearly, the spectrum of the IF signal is the spectrum of the RF signal translated by $\pm\omega_{LO}$. By filtering out the unneeded spectrum portion, the input signal can be either up-converted or down-converted.

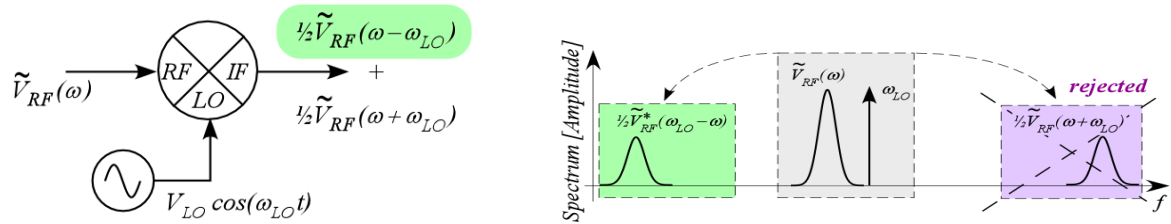

**Fig. 31:** Frequency down-conversion by analog multiplication

In principle any non-linear device could produce the desired frequency translation. If the mixing RF and LO signals are fed into a single diode, under the assumption $V_{LO} \gg V_{RF}$, the LO voltage turns the diode ON and OFF and the IF voltage is

$$V_{IF}(t) = k[V_{LO}(t) + V_{RF}(t)]\cdot[1 - \text{sgn}(V_{LO}(t))] \quad . \tag{25}$$

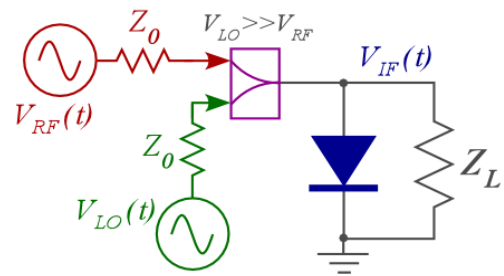

**Fig. 32:** Single diode mixing

With $V_{LO}(t)$ a sine wave, the function $1-\text{sgn}(V_{LO}(t))$ is a square wave expressing the on–off modulation of the diode according to the polarity of the LO voltage. The square wave contains all the odd harmonics of $f_{LO}$, so that each frequency $f_{RF}$ contained in the RF signal produces the output frequencies $f_{IF}$:

$$f_{IF} = n f_{LO} \pm f_{RF} \qquad n = \text{any odd integer} \quad . \tag{26}$$

Owing to the frequency content of the square wave, the real mixer produces many frequency lines other than the $f_{LO} \pm f_{RF}$ ones. These are called 'spurious intermodulation products'.

Actually, real diodes are not ideal switches and on–off commutations are smooth. This effect produces more intermodulation products, so that the frequencies present in the output spectrum are

$$f_{IF} = n f_{LO} \pm m f_{RF} \qquad m, n = \text{any odd integers} \tag{27}$$

Single diode mixing provides no inherent isolation between ports. Lack of isolation results in a large number of intermodulation products, poor conversion loss (i.e., a large value of the ratio between the power of unconverted and converted signals), and various interference and cross-talk problems.

Port isolation is obtained by exploiting symmetries in the mixing network design.

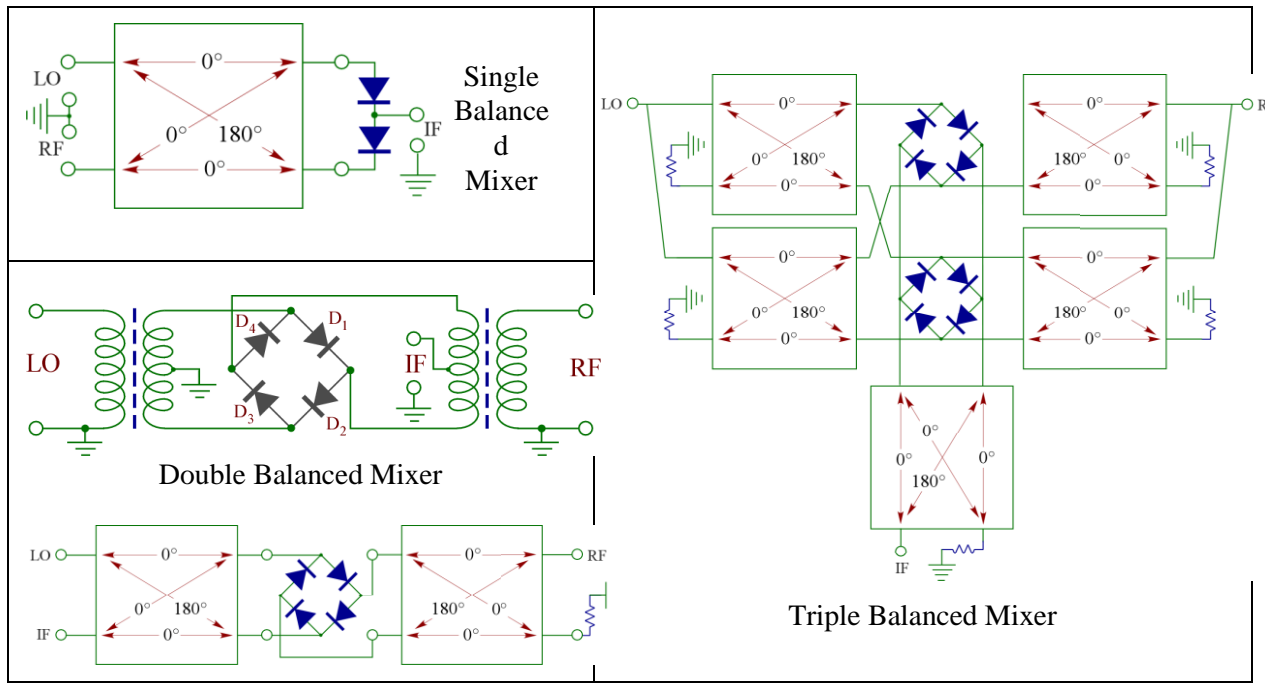

**Fig. 33:** Balanced mixers of various symmetry levels

## 8.1 The double balanced mixer

The double balanced mixer is the most common type of frequency mixer, ensuring good isolation and excellent conversion loss. The LO voltage is differentially applied on the diode bridge switching on/off alternatively the $D_1$–$D_2$ and $D_3$–$D_4$ pairs, so that the IF voltage is given by

$$V_{IF}(t) = V_{RF}(t) \cdot \text{sgn}\left[V_{LO}(t)\right] \quad . \tag{28}$$

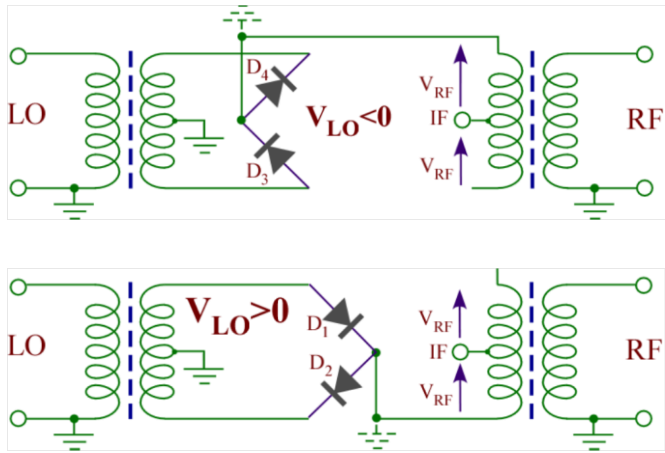 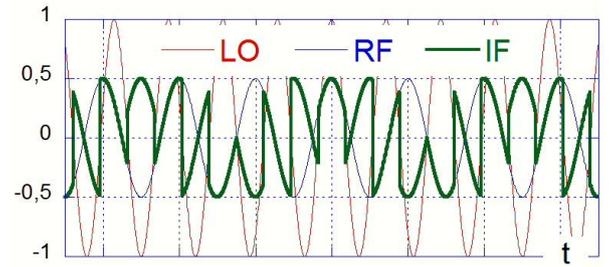

**Fig. 34:** Operating principles of a double balanced mixer

Under the conditions $V_{RF}(t) = A_{RF} \cdot \cos(\omega_{RF} t)$; $V_{LO}(t) = A_{LO} \cdot \cos(\omega_{LO} t)$; $A_{RF} \ll A_{LO}$, Eq. (28) becomes

$$
\begin{aligned}
V_{IF}(t) &= A_{RF} \cos(\omega_{RF} t) \cdot \text{sgn}[\cos(\omega_{LO} t)] = A_{RF} \cos(\omega_{RF} t) \cdot \sum_{n=odds} \frac{4}{n\pi} \cos(n\omega_{LO} t) = \\
&= A_{RF} \cdot \sum_{n=odds} \frac{2}{n\pi} [\cos((n\omega_{LO} - \omega_{RF})t) + \cos((n\omega_{LO} + \omega_{RF})t)] = \\
&= \frac{2}{\pi} A_{RF} [\cos((\omega_{LO} - \omega_{RF})t) + \cos((\omega_{LO} + \omega_{RF})t) + intermod\ products] \quad .
\end{aligned}
\tag{29}
$$

The most relevant specifications qualifying the performance of a double balanced mixer are the following:

– **Frequency range (specific to each port)**
 From DC to > 10 GHz, multi-decades covered by a single device.
 Normally IF band narrower than RF, LO bands. IF may be DC or AC coupled.

– **Mixer level**
 Minimum level at LO to switch on/off the diodes. Typically +3 ÷ +23 dBm, depending on the diode barrier and the number of diodes in series in the bridge.

– **Conversion loss**
 Ratio between the unconverted (RF) and converted (Single Sideband IF) signal levels.
 Theoretical minimum = 3.9 dB ($= 20 \cdot \log[2/\pi]$); practical values in the 4.5 ÷ 9 dB range.
 Is an 'integral' specification. Low CL means also good isolation (not necessarily vice-versa).

– **Isolation**
 Amount of direct signal leakage from one port to another port (reciprocal parameter). L–R critical for interference in the RF circuitry. Typical values 25 ÷ 35 dB.
 L–I critical for filtering when $f_{IF}$ and $f_{LO}$ are close. Typical values 20 ÷ 30 dB.
 R–I is usually not an issue ($P_{RF} \ll P_{LO}$). Typical values 25 ÷ 35 dB.

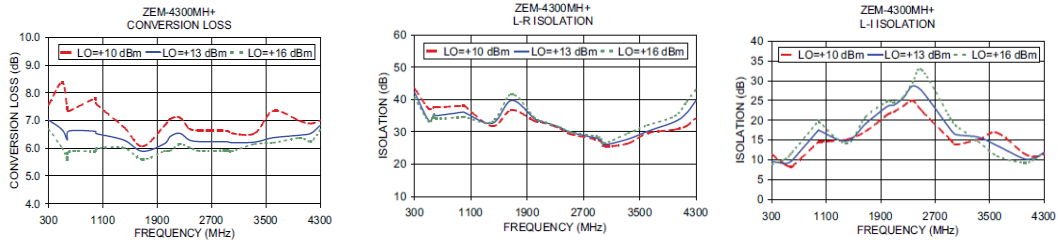

**Fig. 35:** Conversion loss and isolation vs. frequency of a commercial mixer

- **1 dB compression**

    Is a figure of the mixer linearity defined as the RF level showing a 1 dB increase of the conversion loss. Typical values ≈4 dB below mixer specified LO level.

- **Noise figure**

    In a mixer the noise figure is comparable to the conversion loss: $NF_{mix} \geq CL$, $NF_{mix} \approx CL$ (signal reduced, while white noise is unaffected by up/down conversion).
    3 dB worse for SSB (single sideband) w.r.t. DSB (double sideband) signals (signals add coherently, noise quadratically).
    Mixer + IF amp cascades have noise figures $NF = NF_{mix} + CL(NF_{IF} - 1)$. Being magnified by the mixer conversion loss, the IF amp noise figure is crucial.

- **Single and multi-tone intermodulation distortion/two-tone 3$^{rd}$ order intercept**

    Output content of harmonics other than $f_{LO} \pm f_{RF}$. Because of mixer non-linearity, multitone RF signals ($f_{RF_1}$, $f_{RF_2}$, ...) generate output harmonics at $m_1 f_{RF_1} + m_2 f_{RF_2}$. The level of the 3$^{rd}$ order harmonics $2f_{RF_1} - f_{RF_2}$, $2f_{RF_2} - f_{RF_1}$ grows with 3$^{rd}$ power of RF signal, while fundamental $f_{LO} \pm f_{RF}$ tone level grows linearly. Two-tone 3$^{rd}$ order intercept is defined as the RF level where the two output lines cross.

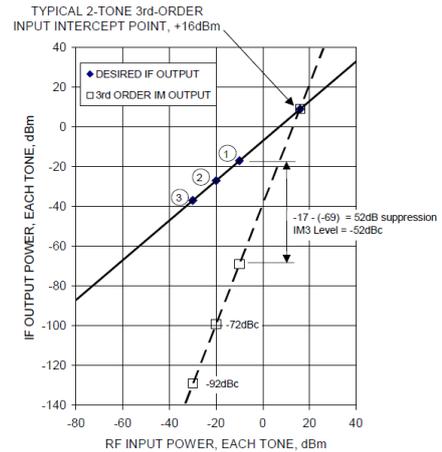

**Fig. 36:** Two-tone 3$^{rd}$ order intercept

Mixers are built in a huge variety of models and packages. They are largely used in RF electronics because they are simple, reliable, passive, and cheap. As will be shown in the following, their applications are not limited to frequency up and down conversion, but they can be used also as phase detectors, bi-phase amplitude modulators, I&Q networks, etc.

## 8.2 Phase detectors (mixer based)

Measuring the phase of RF signals relative to some selected reference is an operation routinely performed in any RF low-level control system in particle accelerators. Beam phase measurement in circular accelerators is an important diagnostic tool to keep the longitudinal dynamics under control. Phase detection can be performed by means of several different techniques. The use of frequency mixers down-converting to baseband the RF signals is the most straightforward approach.

If the signals at RF and LO ports of a mixer are sine waves of equal frequency ($\omega_{LO} = \omega_{RF}$) Eq. (29) shows that the outcoming signal at the IF port has a DC component given by

$$V_{IF}(t) = A_{RF} \cos(\omega t + \phi) \cdot \text{sgn}[A_{LO} \cos(\omega t)] \Rightarrow V_{IF}|_{DC} = \langle V_{IF}(t) \rangle = k_{CL} A_{RF} \cos\phi \qquad (30)$$

where $k_{CL}$ is the mixer conversion loss.

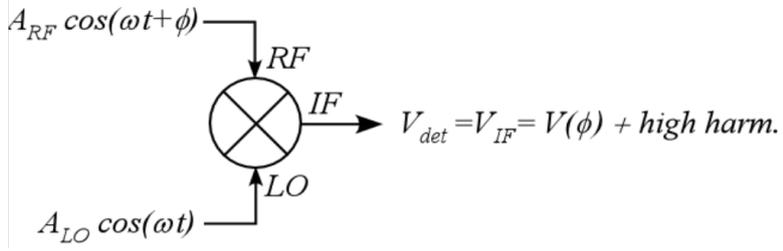
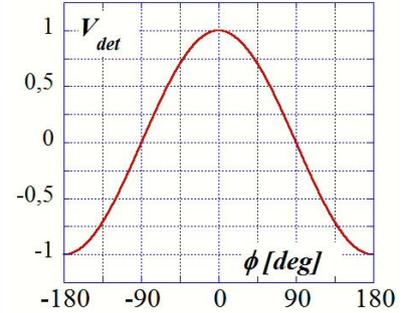

**Fig. 37:** Mixer based phase detector  **Fig. 38:** Detection characteristics

However, mixers dedicated to phase detection can be operated in saturation, i.e., with similar levels at both RF and LO inputs. The diodes are turned on/off by either LO or RF signals, so that Eq. (29) is no longer applicable and the IF signal is given by

$$A_{RF} \approx A_{LO} \Rightarrow V_{IF}(t) = \begin{cases} V_{RF}(t) \cdot \text{sgn}[V_{LO}(t)] & \text{if } |V_{RF}(t)| < |V_{LO}(t)| \\ V_{LO}(t) \cdot \text{sgn}[V_{RF}(t)] & \text{if } |V_{RF}(t)| > |V_{LO}(t)| \end{cases}. \qquad (31)$$

The dc level of the IF voltage can be worked out in this more general excitation configuration. A more linear phase detection characteristics results according to

$$V_{det}(\varphi) = k_{CL} A_{RF} \frac{\sqrt{1 + \alpha^2 + 2\alpha \cos\varphi} - \sqrt{1 + \alpha^2 - 2\alpha \cos\varphi}}{2\alpha} \qquad (32)$$

with $\alpha = \dfrac{A_{RF}}{A_{LO}}$.

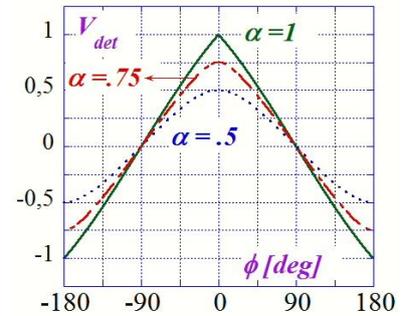

**Fig. 39:** Detection characteristics of a saturated mixer

### 8.3 Phase detectors (other)

Phase detection can be accomplished by analog multiplier circuits or digital comparators operating on squared signals (ex-OR circuits). These detectors show larger sensitivity and output dynamic range, better linearity, but very much smaller bandwidths (typically ≤ 500 MHz).

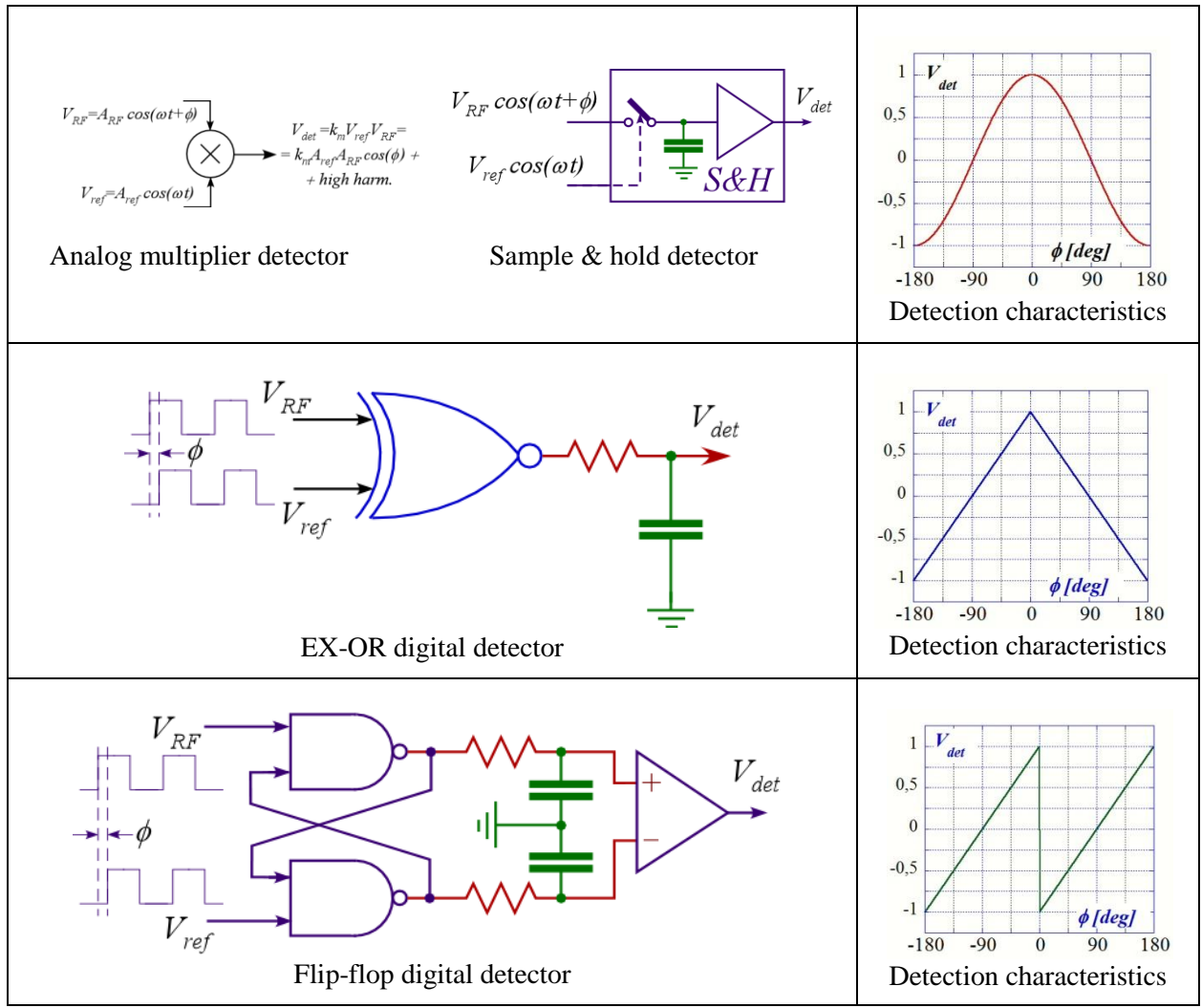

**Fig. 40:** Phase detectors of various types

### 8.4 Bi-phase amplitude modulators

A bi-phase amplitude modulator can be obtained by lowering the L–R isolation of a double balanced mixer in a controlled way by injecting a bias current in the IF port. Positive and negative bias IF currents $I_b$ increase the transconductance of the diode pairs $D_2$–$D_4$ and $D_1$–$D_3$, respectively. According to Eq. (33), the bias current $I_b$ controls the value of the L–R coupling coefficient $k$. The modulator is of the bi-phase type, since the sign of the control current controls the sign of the transfer function.

$$V_{RF}(t) = V_{LO}(t) \cdot \left[ \frac{Z_{D_3}}{Z_{D_3}+Z_{D_4}} - \frac{Z_{D_2}}{Z_{D_1}+Z_{D_2}} \right] = k(I_b) \cdot V_{LO}(t) \quad . \tag{33}$$

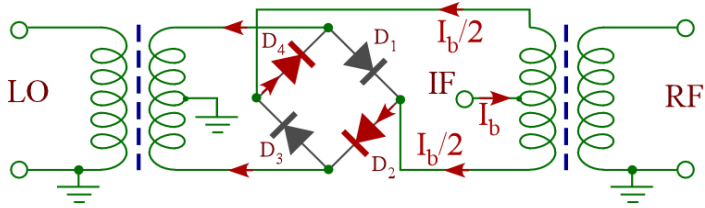
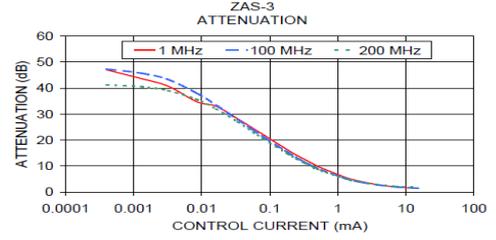

**Fig. 41:** Bi-phase amplitude modulator based on a Double Balanced Mixer (DBM)

**Fig. 42:** Characteristics of a commercial DBM bi-phase amplitude modulator

The device is passive and the coupling coefficient does not depend linearly on the bias current. Moreover, the bandwidth of the IF port can be quite large (DC÷1 GHz typical), which makes this kind of modulator a very fast one.

### 8.5  I&Q networks

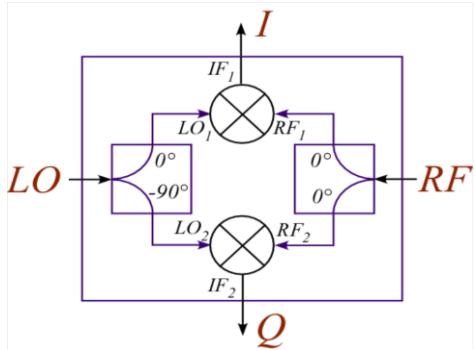

**Fig. 43:** I&Q detector

$$V_I(t) = k_{CL} A_{RF} \cos\left[(\omega_{LO} \pm \omega_{RF})t\right] = A_{IF} \cos(\omega_{IF} t)$$
$$V_Q(t) = k_{CL} A_{RF} \sin\left[(\omega_{LO} \pm \omega_{RF})t\right] = A_{IF} \sin(\omega_{IF} t)$$

if $\omega_{LO} = \omega_{RF}$ (34)

$$\begin{cases} V_I = k_{CL} A_{RF} \cos(\phi) + high\ harmonics \\ V_Q = k_{CL} A_{RF} \sin(\phi) + high\ harmonics \end{cases}$$

The circuit shown in Fig. 43, consisting of two mixers, one splitter and one quadrature hybrid, allows the extraction of in-phase and in-quadrature components of an RF signal. This kind of network is called I&Q detector or I&Q mixer.

In-phase and in-quadrature components of the RF signal are the Cartesian representation of the sine-wave phasor, while amplitude and phase are the equivalent polar representation. The two representations are related by the canonical transformation

$$A_{RF}^2 \div V_I^2 + V_Q^2 \ ; \qquad \tan\varphi = V_Q/V_I \quad . \tag{35}$$

Using mixers as bi-phase controlled attenuators, vector (I&Q) modulators can be obtained in a similar way to I&Q detectors. I and Q copies of the input signal are obtained from a quadrature hybrid and re-combined by a vector combiner after being individually attenuated with independent control signals.

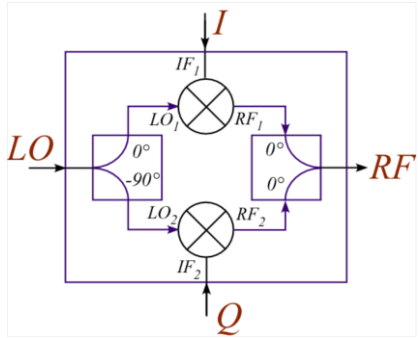

**Fig. 44:** I&Q modulator

$$V_{RF}(t) = A_{LO}\left[k_I \cdot \cos(\omega_{LO} t) + k_Q \cdot \sin(\omega_{LO} t)\right] =$$
$$= A_{LO}\sqrt{k_I^2 + k_Q^2} \cdot \cos\left[\omega_{LO} t + \arctan(k_Q/k_I)\right] \quad . \tag{36}$$

The level of the output signal is controlled by moving $k_I$ and $k_Q$ proportionally, while unbalanced changes produce variations of the output signal phase.

## 8.6 Image rejection mixer

Filtering mixer image frequencies can be difficult and/or costly, especially in up-conversion processes where the image frequency bands are relatively closely spaced. By cascading an I&Q mixer and a quadrature hybrid an 'image rejection' network is obtained, where the two image signals are separately available at two physically different ports. No narrowband filtering is then necessary to separate the two output signal components.

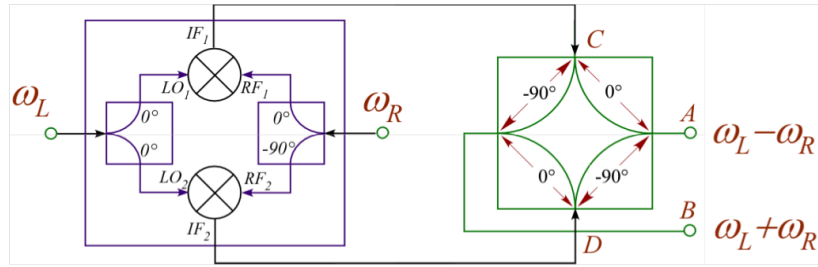

**Fig. 45:** Image rejection mixer (this configuration requires $\omega_L > \omega_R$ to work properly)

The IF signals generated by the I&Q mixers are given by

$$V_{IF_1}(t) = k_{CL} A_{RF} \left[ \cos\left((\omega_L + \omega_R)t\right) + \cos\left((\omega_L - \omega_R)t\right) \right] ,$$
$$V_{IF_2}(t) = k_{CL} A_{RF} \left[ \sin\left((\omega_L + \omega_R)t\right) - \sin\left((\omega_L - \omega_R)t\right) \right] ,$$
(37)

which combine in the quadrature hybrid according to the scattering matrix of Fig. 12, so that each of the two original frequencies survive at only one of the two output ports while it cancels out at the other.

$$V_A(t) = k\, A_{RF} \cos\left((\omega_L - \omega_R)t\right) \;;\; V_B(t) = k\, A_{RF} \sin\left((\omega_L + \omega_R)t\right) .$$
(38)

## 9    Peak detectors

Diode **peak detectors** are used to **sample** the amplitude of RF signals. They basically work as rectifiers, sampling the RF peak while charging a capacitance, and holding the peak voltage, slowly discharging the capacitance on a load (typically 50 Ω to follow fast level variations). A schematics of a peak detector circuitry is shown in Fig. 46, while the detector saw-tooth output voltage following the RF level envelope is shown in Fig. 47.

**Schottky diodes** are used for zero-bias, very **broadband sensors** (up to 50 GHz).

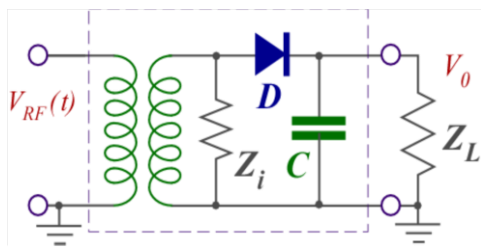

**Fig. 46:** Peak detector schematics

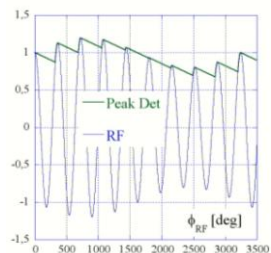

**Fig. 47:** Detector voltage following RF peaks

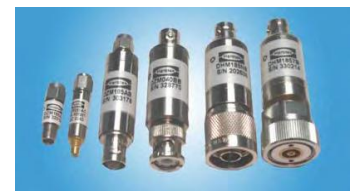

**Fig. 48:** Commercial peak detectors of various shapes

A typical detection curve is shown in Fig. 49. For low levels of the input RF, the detector output is proportional to the square of the input amplitudes, i.e., the output voltage is proportional to the input power. In the region where the input level approaches 1 V the characteristic is pretty linear with a negative offset term. At larger input levels the device response linearity is worsened by the saturation.

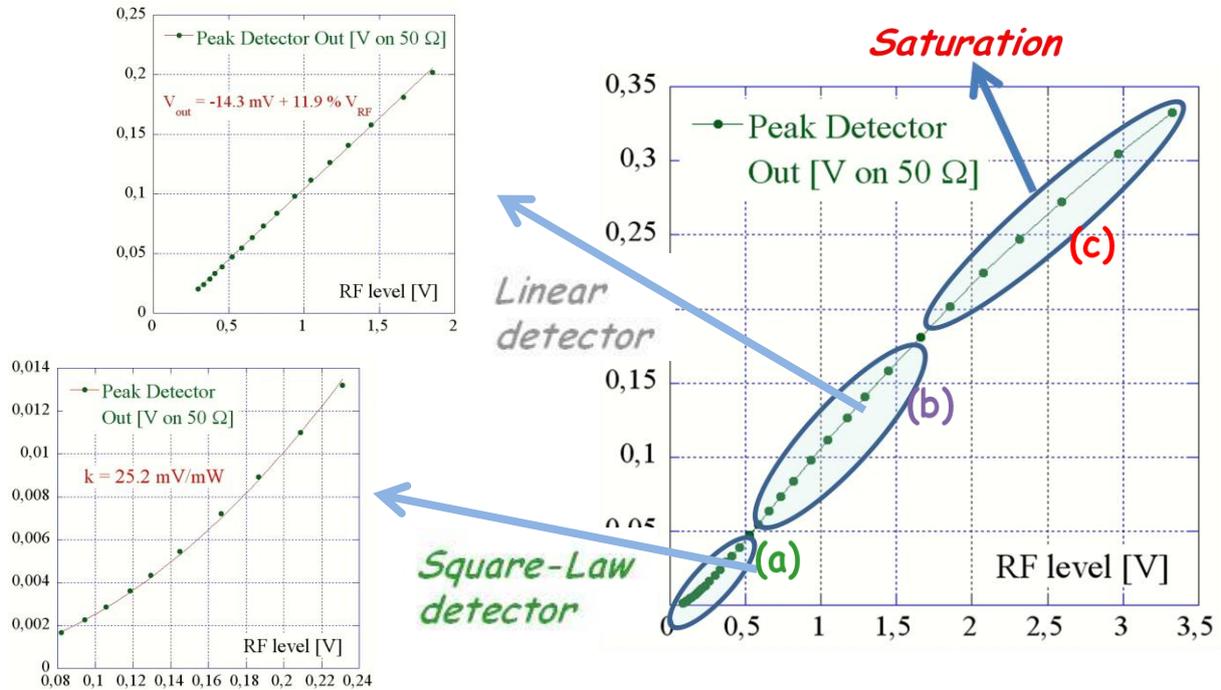

**Fig. 49:** Typical detection curve showing three regions: square-law (a), linear (b), and saturation (c)

## 10  Step recovery diodes

Diodes switching from forward to reverse polarization deliver in a certain time all the charge stored on both sides of the space-charge/depletion region. Step Recovery Diodes (SRDs) have a PIN-like structure with a special doping profile allowing the reverse current to circulate across the depletion region for a short time before abruptly dropping to zero in a few tens of ps.

The sharp variation of the circuit current can be used to generate narrow voltage pulses on a load. The spectrum of the output signal is a series of peaks containing all the harmonics of the input sine-wave and may extend well beyond 10 GHz.

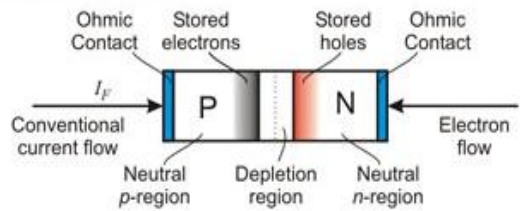

**Fig. 50:** P-I-N structure of a step recovery diode

SRDs are used to generate a train of very short pulses in time domain corresponding to a spectrum of regularly spaced lines (comb generator). Any required harmonics of the input signal can be extracted by properly filtering the output voltage, so that this kind of device can also be used as a frequency multiplier.

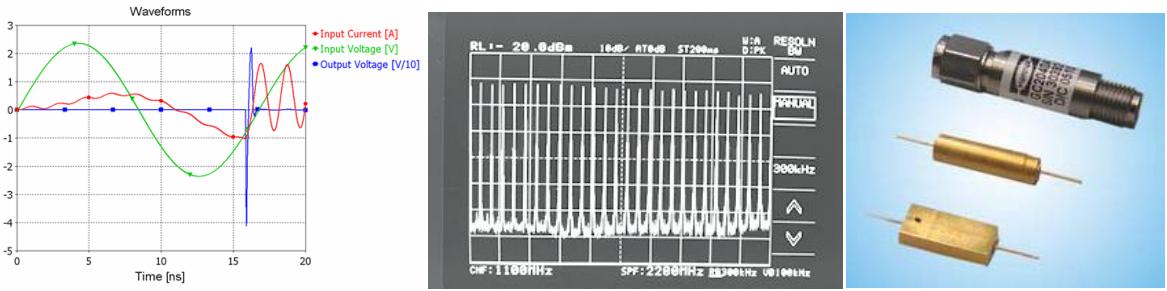

**Fig. 51:** Step recovery diode: (a) time domain pulse; (b) output spectrum; (c) commercial models

## 11 PIN diodes – variable attenuator/switch

The dynamic impedance shown by biased PIN diodes for RF applications is inversely proportional to the bias current. This makes PIN diodes suitable controlled resistors to be put in T or Π configurations of resistive attenuators.

The advantage of PIN technology relies on the fact that the intrinsic layer resistance remains dominant in forward bias, while in standard PN diodes the junction diffusion capacitance shorts the device at high frequencies.

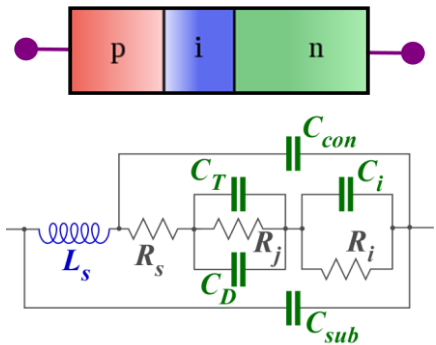
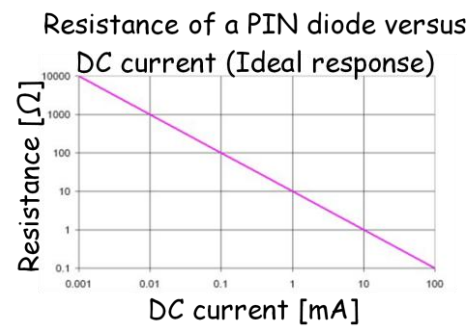

**Fig. 52:** Equivalent RF circuit of a PIN diode    **Fig. 53:** PIN diode resistance vs. bias current

The possibility of controlling the diode resistance through the bias current is exploited in electronic variable attenuators based on PIN diode technology. The attenuator architecture complexity depends on the required performance, especially concerning the matching of the RF ports. A single-cell attenuator, a simple structure depicted in Fig. 54, works as a reflective device since the diode variable resistance value affects mainly the input reflection coefficient. Matching of such a network is quite poor. A λ/4 diode cascade shows better performance in terms of dynamic range and input matching (see Fig. 55). Optimal results are obtained with the diodes unequally biased in a proper way. Reflective attenuators can be matched by inserting a 90° hybrid as shown in Fig. 56 provided that the two diodes are a pair equally biased. The two reflected waves cancel out at the input port and add-up at the output.

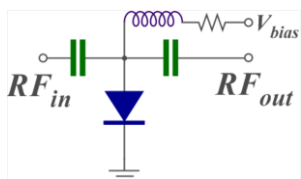
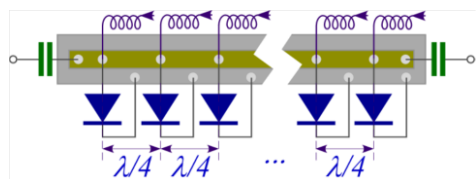
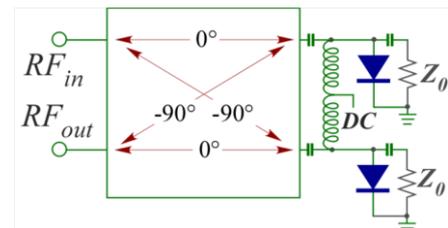

**Fig. 54:** Single-cell attenuator    **Fig. 55:** Multi-cell, λ/4 cascade attenuator    **Fig. 56:** 90° hybrid balanced attenuator

In the extreme bias conditions the diodes are on/off (=open/short) so that an RF signal can be fully transmitted or fully stopped and the device acts as a controlled RF switch.

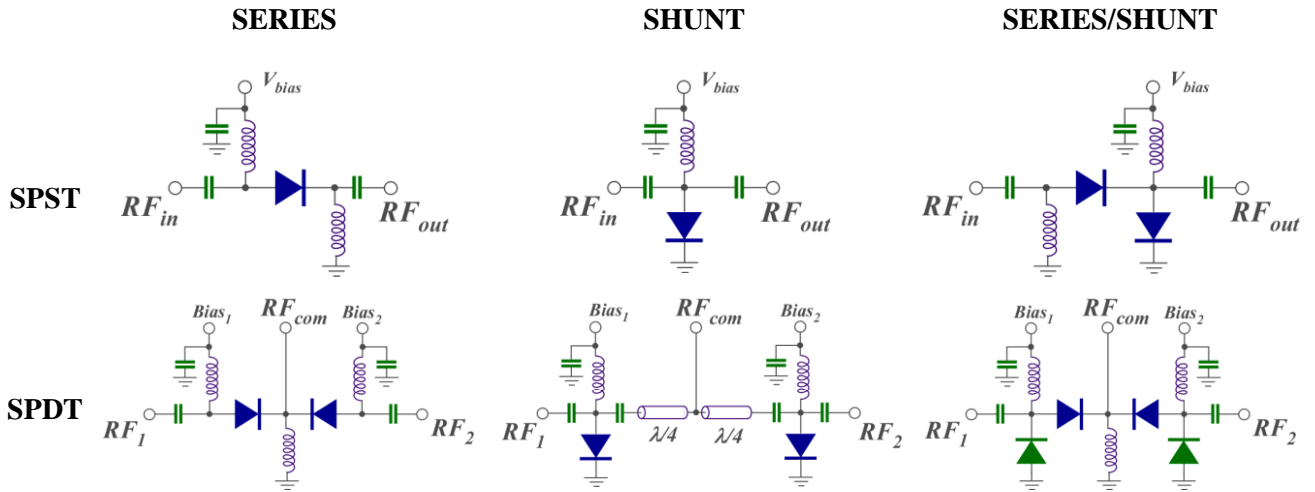

**Fig. 57:** PIN diode RF switches of various types and topologies

Single-Pole Single-Through (SPST) and Single-Pole Double-Through (SPDT) electronic switches make use of PIN diodes in various configurations. Series circuits suffer from poor isolation since transition capacitance of reversely biased diodes allows significant RF signal transmission.

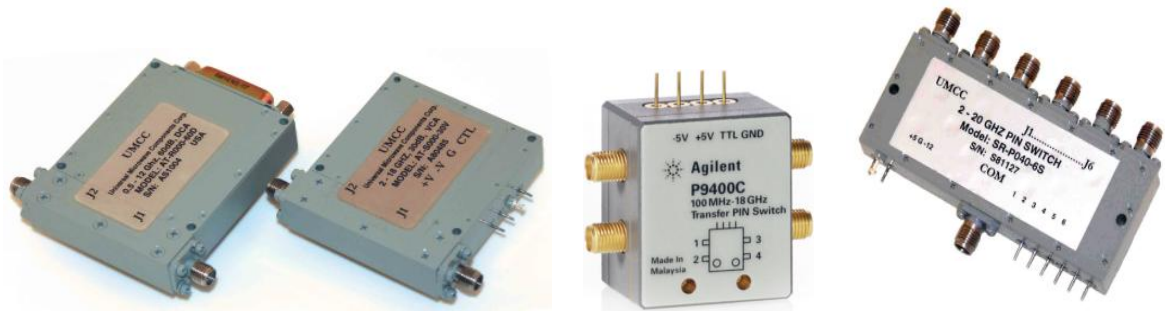

**Fig. 58:** Commercial PIN diode switches of various types

Shunt circuits perform better in terms of isolation, but device matching is poor and the presence of λ/4 lines in the SPDT version limits the device bandwidth. Compound series/shunt circuits offer the best performance, although a large number of diodes increases the device insertion loss.

The most relevant specifications of electronic attenuators and switches are listed in the following

*Attenuators*

- **Frequency range**

  From DC to > 10 GHz, multi-octaves

- **Level**

  Maximum power at the input

  (typ. 10÷30 dBm)

- **Insertion loss**

  Minimum device attenuation (typ. 1÷6 dB)

*Switches*

- **Frequency range**

  From DC to > 10 GHz, multi-decades

- **Level**

  Maximum power at the input

  (typ. 10÷30 dBm)

- **Insertion loss**

  Attenuation in the 'ON' state (typ. 1÷3 dB)

- **Isolation**

  Signal transmission at maximum attenuation (typ. 30÷80 dB)

- **Dynamic range**

  Excursion of the available attenuation values (typ. 30÷80 dB)

- **Flatness**

  Attenuation fluctuation over the frequency range at fixed control voltage (typ. 1÷3 dB)

- **Control bandwidth**

  Modulation frequency producing a peak AM 3 dB lower than that produced by a low-frequency voltage of the same value

- **Isolation**

  Signal transmission in the 'OFF' state to the output (SPST) or to the unselected port (SPDT) (typ. 25÷80 dB)

- **Switching time**

  Minimum time required to turn the device ON/OFF (typ. > 5 ns)

## 12  Variable delay lines/phase shifters

Phase shifters are devices ideally capable of transmitting an RF signal shifting its phase to any desired value without attenuation. Depending on the nature of the control mechanism, phase shifters can be classified as mechanical or electrical, and continuously or digitally (i.e., in steps) variable.

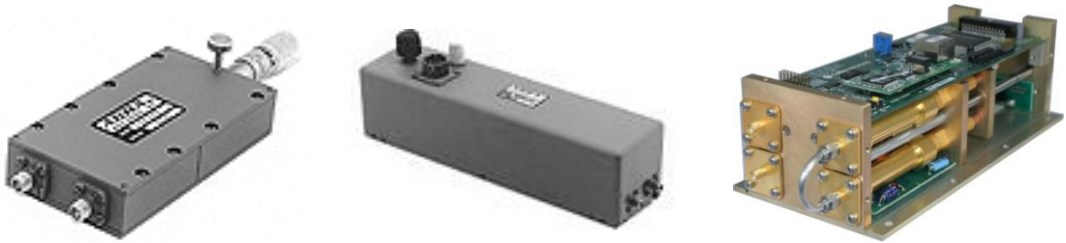

**Fig. 59:** Commercial manual and motorized 'trombones'

Some pictures of commercially available stretched delay lines (trombones) are shown in Fig. 59. These devices are mechanical, continuously variable, low attenuation and very broadband, and may be used whenever variation speed is not an issue. However, they are in general expensive and not very compact if compared to other solutions.

Much cheaper and faster phase shifters are based on a 90° hybrid junction loaded by biased varactor diodes. The RF signal sees the transition capacitance of the matched, inverse biased varactor diode pair which depends on the control voltage.

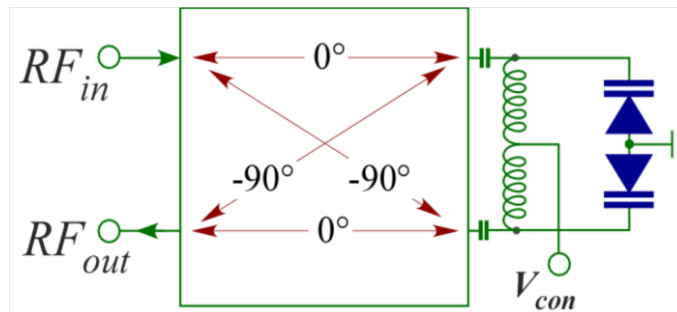

**Fig. 60:** Electronic phase shifter based on a 90° hybrid loaded by varactor diodes

Ideally, the scattering matrix of this device is perfectly matched since reflected waves from the varactor pair cancel out at the input port so that $s_{11} = s_{22} = 0$, while they add-up at the output port. The varactor reflection is ideally full (fully reactive load), and the reflection phase depends on the varactor transition capacitance (tunable via the bias) and on the excitation frequency, according to Eq. (39).

$$S = \begin{pmatrix} s_{11} & s_{12} \\ s_{21} & s_{22} \end{pmatrix} = \begin{pmatrix} 0 & e^{j\Delta\varphi_{out-in}} \\ e^{j\Delta\varphi_{out-in}} & 0 \end{pmatrix}$$

$$s_{11} = s_{22} = \frac{V_{ref_{in}}}{V_{fwd_{in}}} = \frac{1}{2}\left[\frac{1-j\omega C_T(V_{con})Z_0}{1+j\omega C_T(V_{con})Z_0} + (-j)^2 \frac{1-j\omega C_T(V_{con})Z_0}{1+j\omega C_T(V_{con})Z_0}\right] \approx 0;$$

$$s_{21} = s_{12} = \frac{V_{fwd_{out}}}{V_{fwd_{in}}} = -j\frac{1-j\omega C_T(V_{con})Z_0}{1+j\omega C_T(V_{con})Z_0} \qquad |s_{21}| = 1$$
(39)

$$\Delta\varphi_{out-in} = -\frac{\pi}{2} - 2\arctan\left[\omega C_T(V_{con})Z_0\right] = 2\arctan\left[\omega_0(V_{con})/\omega\right] + \frac{\pi}{2} \quad \text{with} \quad \omega_0(V_{con}) = 1/(C_T Z_0) \quad .$$

Analog continuously variable phase shifters are typically narrowband ($f_{BW}$ of the order 10 % of $f_0$) but available in a wide frequency range extending to $\approx 10$ GHz. The dependence of the phase insertion on the control voltage is in general non-linear. A certain amount of insertion loss is also present (1÷4 dB typically) which may vary slightly with the bias voltage. The control bandwidth can extend beyond 1 MHz, so that this kind of shifter can also be effectively used as phase modulator.

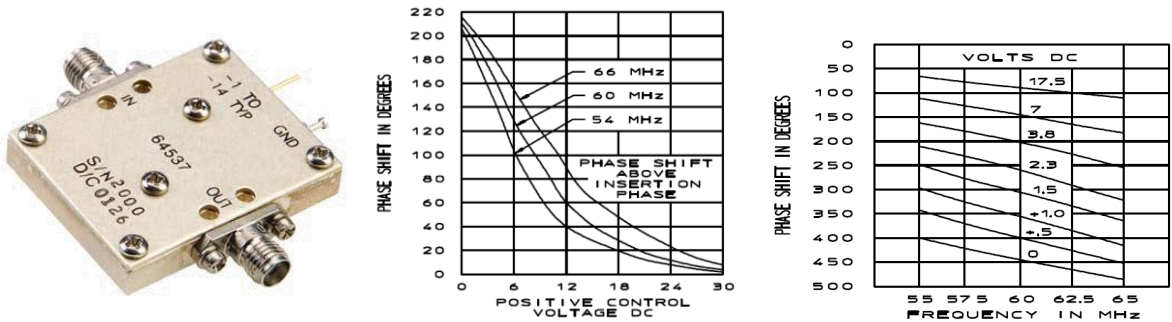

**Fig. 61:** A commercial phase shifter model together with its characteristics

Digital phase shifters are based on PIN diode arrays. A digital control word sets the status of the array of RF switches so that the input signal is routed along paths of different electrical lengths. Depending on the structure of the basic cell, digital phase shifters are of the following types:

– Periodically loaded line
– Switched-line
– Hybrid-coupled line

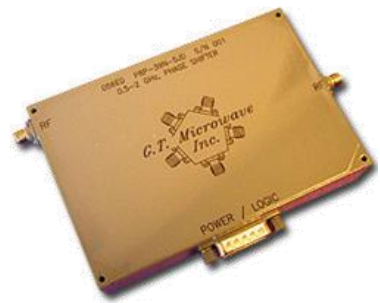

**Fig. 62:** A commercial digital phase shifter model

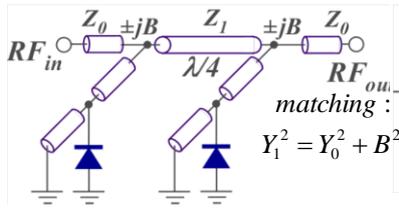
**Fig. 63:** Periodically loaded line phase-shifter

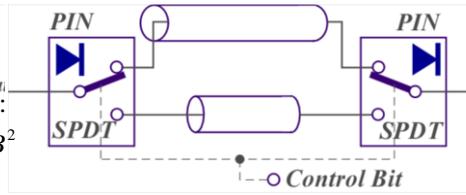
**Fig. 64:** Switched-line phase shifter

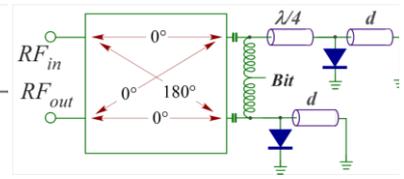
**Fig. 65:** Hybrid-coupled line phase shifter

## 13 Voltage controlled oscillators (VCOs)

Voltage Controlled Oscillators (VCOs) are RF oscillators whose actual output frequency can be controlled by the voltage present at a control (tuning) port.

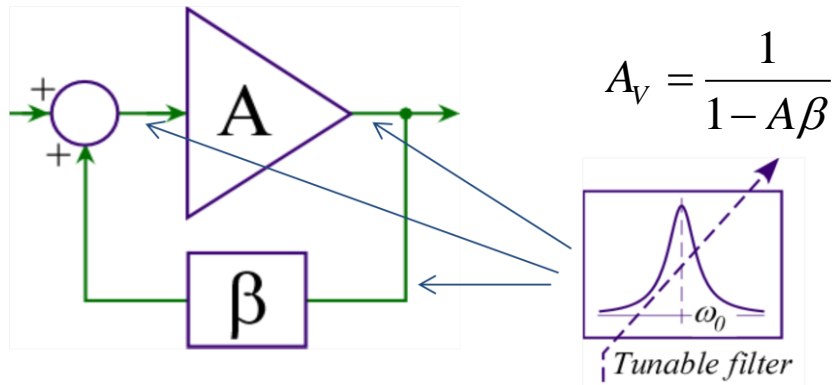
**Fig. 66:** Oscillators are unstable feedback systems

Oscillators are linear feedback systems where stability conditions are not fulfilled. According to the Barkhausen criterion a feedback systems breaks into oscillations at frequencies where the loop gain $G = A\beta$ is such that

$$|G(j\omega)| = 1 \quad ; \quad \angle G(j\omega) = 2n\pi \quad . \tag{40}$$

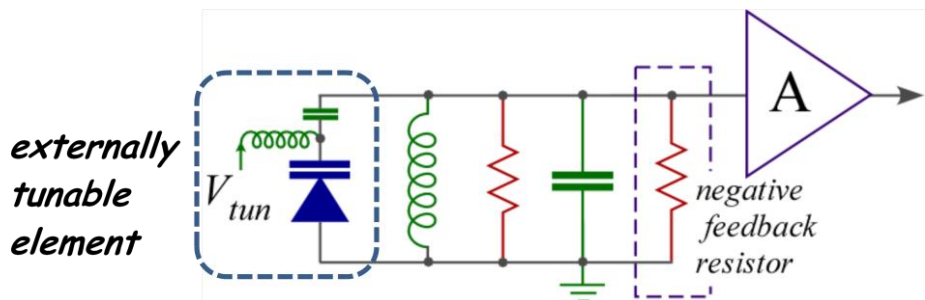
**Fig. 67:** Oscillator lossless resonant filter model

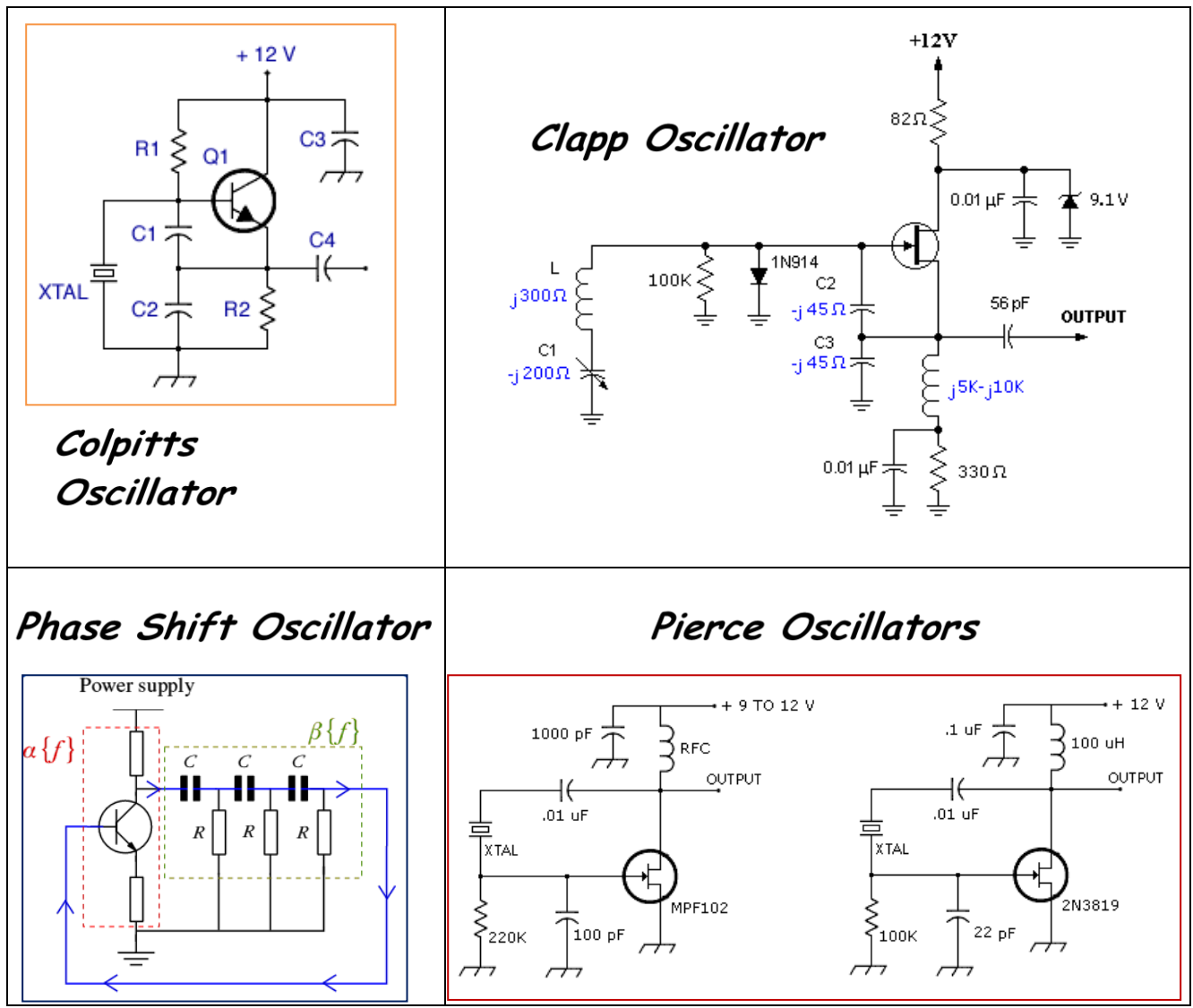

**Fig. 68:** Oscillators of various types

More realistically, oscillation occurs at frequencies where the small signal linear gain $G_s$ is slightly in excess of 1:

$$|G_s(j\omega)| > 1 \quad ; \quad \angle G_s(j\omega) = 2n\pi \tag{41}$$

and are confined at amplitudes where non-linearity (compression) sets the large signal $G_L$ at $|G_L(j\omega)| \approx 1$. Oscillation frequency is controlled by inserting tunable filters in the loop, so that the loop gain can bump beyond the limit value 1 only within the filter frequency band. Positive feedback can be modelled as a negative resistor compensating the losses of the filter, which behaves as a lossless resonator.

There are a number of possible oscillator architectures. VCOs use varactors as tuning elements. Resonant tuning filters can be lumped, transmission-line-based, or dielectric resonators (DROs). The most important specifications that qualify a VCO for RF applications are the following:

- **Tuning characteristics**
  Frequency versus tuning voltage plot.

- **Tuning sensitivity**
  Slope of the tuning characteristics, typically given in MHz/V. Is a local parameter in case the tuning characteristic is not linear over the entire range.

- **Temperature sensitivity**
  Frequency variation with temperature at a fixed tuning voltage.

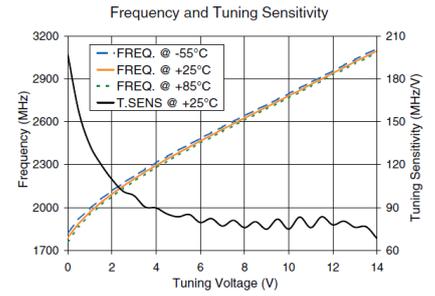

- **Modulation bandwidth/Tuning speed**
  Modulation frequency producing a peak frequency deviation reduced by 3 dB compared to that produced by a dc voltage of the same value/Time required to settle the output frequency deviation to 90% of the regime value after application of a voltage step variation on the tuning port. The two parameters are obviously correlated.

- **Output power/Output power flatness**
  Level of the oscillator output fundamental harmonic into a 50 Ω load/Variation of the output level over the specified VCO frequency range.

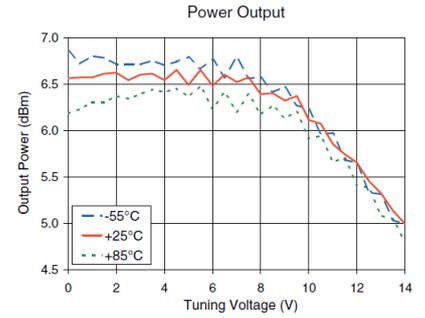

- **Frequency pushing/Frequency pulling**
  Variation of the VCO frequency with the supply voltage at fixed control voltage/Variation of the output frequency with the load mismatch (typically given as peak-to-peak value at 12 dB return loss, any phase).

- **Harmonic suppression**
  Level of the harmonics relative to the fundamental (typically given in dBc = dB below the carrier).

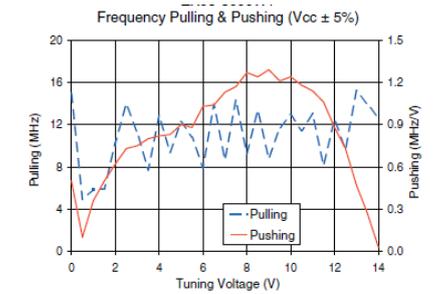

- **Spurious content**
  Level of the spurious, non-harmonic output signals relative to the oscillator output (typically given in dBc).

- **SSB phase noise**
  Single-sideband phase noise in 1 Hz bandwidth as a function of the frequency offset from the carrier frequency, measured relative to the carrier power and given in dBc/Hz. Very important to evaluate the expected residual phase noise in phase locked loops.

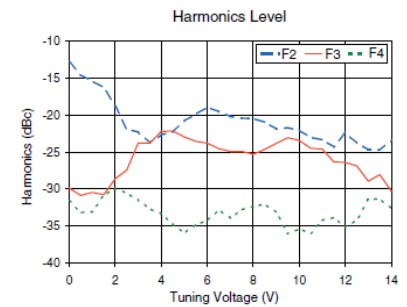

- **r.m.s. phase jitter**
  r.m.s. value of the instantaneous phase deviation, which is given by the integral of the SSB power spectrum:

$$\phi_{rms}^2 = \frac{1}{\Delta T} \int_{t_0}^{t_0+\Delta T} \phi^2(t)\,dt = 2\int_{f_L}^{f_H} 10^{-(SSB_{dBc}/10)}\,df \quad .$$

r.m.s. jitter expressed in terms of frequency deviation is known as '*residual FM*', defined as the SSB power spectrum integral between $f_L = 50$ Hz and $f_H = 3$ kHz:

$$\Delta f_{rms}^2 = \frac{1}{\Delta T} \int_{t_0}^{t_0+\Delta T} \Delta f^2(t)\,dt = 2\int_{f_L}^{f_H} 10^{-(SSB_{dBc}/10)} f^2\,df \quad .$$

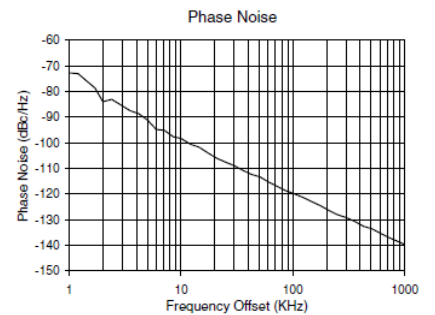

**Fig. 69:**
Characteristic plots of a real VCO

## 14  Phase locked loops (PLLs)

Phase Locked Loops (PLLs) are a very general subject in RF electronics. They are used to synchronize oscillators to a common reference or to extract the carrier from a modulated signal (FM tuning). Therefore they are widely used in receivers, in converters, and in any kind of synchronization hardware present in particle accelerators. The PLL main components are:

- a VCO, whose frequency range includes $Nf_{ref}$;
- a phase detector, to compare the scaled VCO phase to the reference;
- a loop filter, which sets the lock bandwidth;
- a prescaler (by-N frequency divider), which allows setting different output frequencies w.r.t. the reference one.

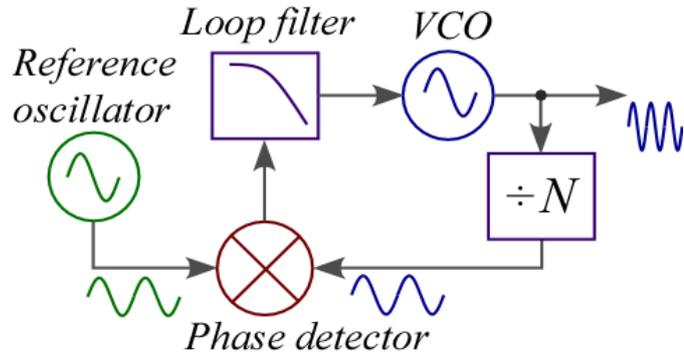

**Fig. 70:** Block diagram of a PLL

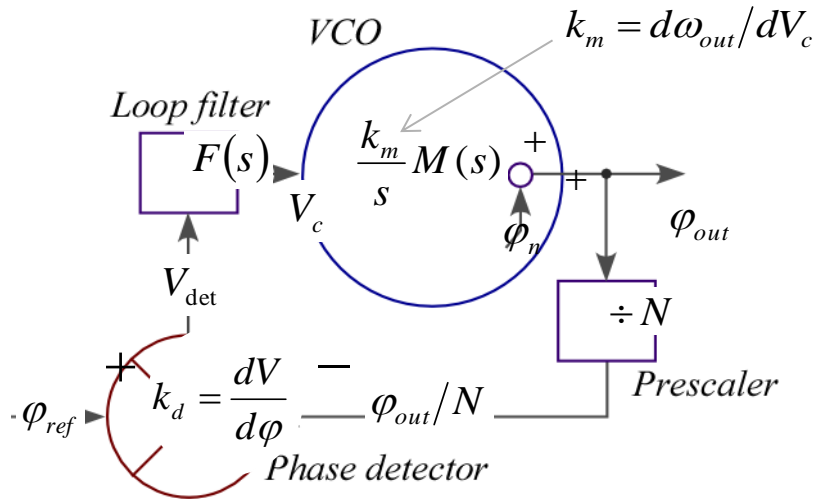

**Fig. 71:** PLL linear model

A PLL linear model including functional transfer function of each constituting block is depicted in Fig. 71.

The transfer function relating the phase of the output signal to both the phase of the reference signal and the VCO phase noise is calculated in Eq. (42):

$$\phi_{out}(s) = N\frac{H(s)}{1+H(s)}\phi_{ref}(s) + \frac{1}{1+H(s)}\phi_n(s) \qquad \text{with } H(s) = \frac{k_d k_m}{N s}F(s)M(s) \;, \tag{42}$$

where $F(s)$ and $M(s)$ are the transfer functions describing the loop filter and the VCO modulation response, while an integrator term $1/s$ results in the loop transfer function $H(s)$ because of the frequency-to-phase conversion.

Loop filters provide PLL stability, tailoring the frequency response, and set loop gain and cut-off frequency.

The phase spectrum of a PLL system together with those of the reference oscillator and of the free-run VCO measured in a test bench is shown in Fig. 72. According to Eq. (42), the output phase spectrum is locked to the reference at frequencies where $|H(j\omega)| \gg 1$, while it returns like the free-run VCO whenever $|H(j\omega)| \ll 1$.

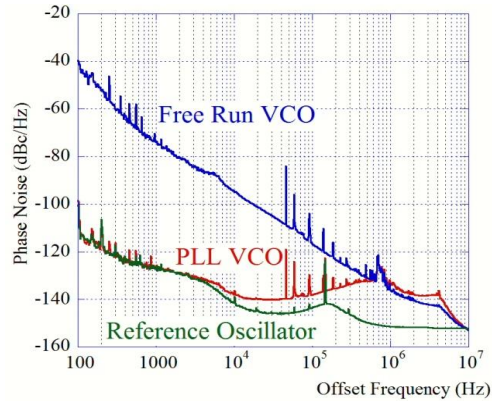

**Fig. 72:** Measured phase noise spectra of a PLL system (including reference oscillator and free-run VCO)

A flat-frequency response loop filter gives a pure integrator loop transfer function thanks to a pole in the origin ($f = 0$) provided by the dc frequency control of the VCO. However, the low-frequency gain can be further increased with a loop filter providing an extra pole in the origin and a compensating zero at some non-zero frequency ($\omega_{zero} = 1/R_2 C$). A very steep loop frequency response is obtained (slope = 40 dB/decade) in stability conditions, as shown in the loop gain Bode and Nyquist plots reported in Fig. 74.

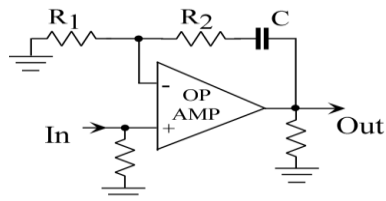

**Fig. 73:** Loop filter providing an extra pole in the origin

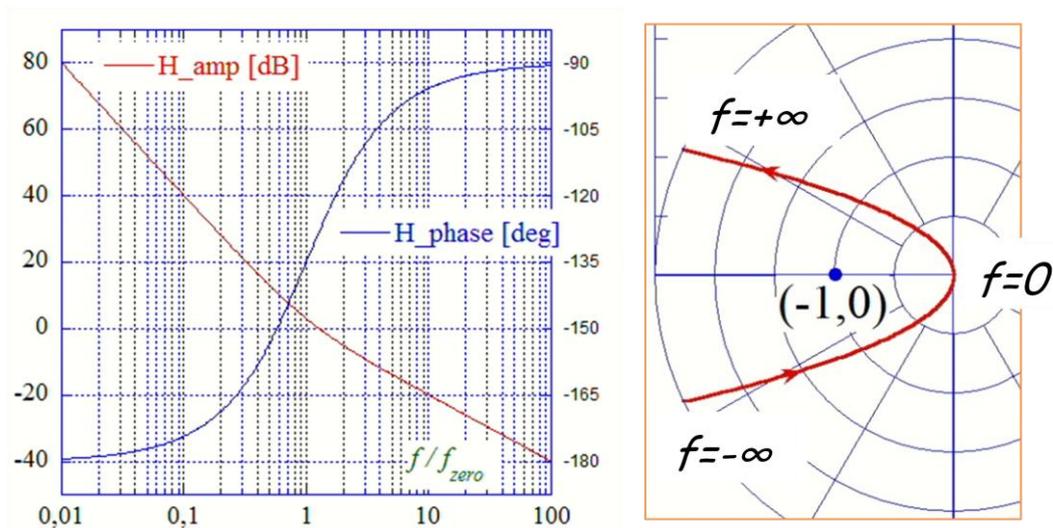

**Fig. 74:** Loop gain Bode and Nyquist plots of a PLL system with two poles in the origin and 1 zero

**Bibliography**

- R. Garoby, *Low-level RF building blocks*, CAS-CERN Accelerator School: RF Engineering for Particle Accelerators, Oxford, 1991, CERN-92-03 vol. 2, p. 428.
- F. Caspers, *Basic concepts II*, CAS-CERN Accelerator School: RF Engineering for Particle Accelerators, Oxford, 1991, CERN-92-03 vol. 1, p. 125.
- H. Henke, *Basic concepts I and II,* CAS-CERN Accelerator School: Radio Frequency Engineering, Seeheim, Germany, 2000, CERN-2005-003, p. 65.
- P. Baudrenghien, *Low-level RF systems for synchrotrons. Part II: High intensity. Compensation of beam-induced effects*, CAS-CERN Accelerator School: Radio Frequency Engineering, Seeheim, Germany, 2000, CERN-2005-003, p. 175.
- R. E. Collin, *Foundations for Microwave Engineering*, 2$^{nd}$ ed. (Mc Graw-Hill, New York, 1992).
- S. Ramo, J. R. Winery, and T. Van Duzer, *Fields and Waves in Communication Electronics*, 3$^{rd}$ ed. (Wiley, New York, 1994).
- J. Millman, *Microelectronics: Digital and Analog Circuits and Systems* (Mc Graw-Hill, New York, 1979).
- H. Taub and D. L. Schilling, *Principles of Communication Systems* (Mc Graw-Hill, New York, 1971).
- G. Kennedy and R. W. Tinnell, *Electronic Communication Systems* (Mc Graw-Hill, New York, 1970).
- U. L. Rhode, *Digital PLL Frequency Synthesizers* (Prentice Hall, Englewood Cliffs, NJ, 1983).
- J. Gorski-Popiel, *Frequency Synthesis: Techniques and Applications* (IEEE Press, Piscataway, NJ, 1975).
- F. M. Gardner, *Phaselock Techniques* (Wiley, Newark, NJ, 2005) e-book.
- S. D'Agostino and S. Pisa, *Sistemi elettronici per le microonde*, Masson editoriale ESA.
- Microsemi-Watertown, The PIN diode circuit designers' handbook,

    http://www.microsemi.com/brochures/pindiodes/page1.pdf
- MINI-CIRCUITS Application Notes, http://minicircuits.com/pages/app_notes.html
- MERRIMAC Application Notes, http://www.merrimacind.com/rfmw/appnotes.html
- http://www.rfcafe.com/references/app-notes.htm